\documentclass[twocolumn,showpacs,aps,pra]{revtex4}
\usepackage{amssymb}
\usepackage{amsfonts}
\usepackage{dcolumn}
\usepackage{amsmath}
\usepackage{graphicx}
\usepackage{psfrag}
\usepackage{amsmath}
\usepackage{amssymb}
\usepackage{epstopdf}
\usepackage{bbold}
\newcommand{\be} {\begin{equation}}
\newcommand{\ee} {\end{equation}}
\newcommand{\nbar} {\bar{n}}

\begin{document}

\title{Quantum theory of a bandpass Purcell filter for qubit readout}
\author{Eyob A. Sete$^1$\footnote{Present address: Rigetti Quantum Computing, 2855 Telegraph Ave, Berkeley, CA 94705, USA}, John M. Martinis$^{2,3}$, and Alexander N. Korotkov$^1$}
\affiliation{$^1$Department  of Electrical and Computer Engineering,
University of California, Riverside, California 92521, USA\\
$^2$Department of Physics, University of California, Santa Barbara,
California 93106, USA \\
$^3$Google Inc., Santa Barbara, USA}
\date{\today}

\begin{abstract}
The measurement fidelity of superconducting transmon and Xmon qubits is
partially limited by the qubit energy relaxation through the resonator into
the transmission line, which is also known as the Purcell effect. One
way to suppress this energy relaxation is to employ a filter which impedes
microwave propagation at the qubit frequency. We present
semiclassical and quantum analyses for the bandpass Purcell filter
realized by E.\ Jeffrey \textit{et al}.\ [Phys.\ Rev.\ Lett.\ 112,
190504 (2014)]. For typical experimental parameters, the bandpass
filter suppresses the qubit relaxation rate by up to two orders of
magnitude while maintaining the same measurement rate. We also show
that in the presence of a microwave drive the qubit relaxation rate
further decreases with increasing drive strength.
\end{abstract}
\pacs{03.67.Lx,  85.25.-j, 03.65.Yz}

\maketitle
\section{Introduction}

The implementation of  fault-tolerant quantum information processing
\cite{N-C-book} requires high-fidelity quantum gates and also needs sufficiently fast and accurate qubit measurement. Superconducting quantum computing technology \cite{Barends-14,Chow-14,Weber-14,Sun-14,Riste-14,Stern-14,Mlynek-14,Lin-14,Kelly-15} is currently approaching the threshold for quantum error correction. Compared with the recent rapid progress in the increase of single-qubit and two-qubit gate fidelities, qubit measurement shows somewhat slower progress. The development of faster and higher-fidelity qubit readout remains an important task.

In circuit QED (cQED) \cite{Blais-04,Wallraff-04}, the qubit state
is inferred by measuring the state-dependent frequency shift of the
resonator via homodyne detection. This method introduces an unwanted
decay channel \cite{Esteve-86} for the qubit due to the energy
leakage through the resonator into the transmission line, the
process known as the Purcell effect \cite{Purcell-46,Haroche-book}.
The Purcell effect is one of the limiting factors for high fidelity
qubit readout.

In principle, the Purcell rate can be suppressed by increasing the
qubit-resonator detuning, decreasing the qubit-resonator coupling,
or decreasing the resonator bandwidth due to damping. However,
these simple methods increase the time needed to measure the qubit.
This leads to a trade-off between the qubit relaxation and
measurement time, whereas it is desirable to suppress the Purcell rate
without compromising qubit measurement. Several proposals have been
put forward for this purpose, which include employing a Purcell
filter \cite{Red10,Jeffrey-14,Bronn-APS-15,Bronn-15}, engineering a Purcell-protected
qubit \cite{Gam11,Srinivasan-11}, or using a tunable coupler that
decouples the transmission line from the resonator during the
qubit-resonator interaction, thereby avoiding the Purcell effect
altogether \cite{Set13}.

The general idea of the Purcell filter is to impede the propagation of
the photon emitted at the qubit frequency, compared with propagation
of the microwave field at the resonator frequency, used for the
qubit measurement. A notch (band-rejection) filter detuned by 1.7
GHz from the resonator frequency was realized in Ref.\ \cite{Red10}.
A factor of 50 reduction in the Purcell rate was demonstrated when
the qubit frequency was placed in the rejection band of the filter.
A bandpass filter with the quality factor $Q_{\rm f}\simeq 30$ (and
corresponding bandwidth of 0.22 GHz) centered near the
resonator frequency was used in Ref.\ \cite{Jeffrey-14}. This
allowed the qubit measurement within 140 ns with fidelities
$F_{|1\rangle}=98.7$ and $F_{|0\rangle}=99.3$ for the two qubit
states. (The bandpass Purcell filter was also used in Ref.\ \cite{Kelly-15}; it
had a similar design with a few minor changes.) A major advantage of
the bandpass Purcell filter in comparison with the notch filter is
the possibility to keep strongly reduced Purcell rate for qubits
with practically any frequency (except near the filter frequency),
thus allowing quantum gates based on tuning the qubit frequency,
and also allowing multiplexed readout of several qubits by placing
readout resonators with different frequencies within the filter
bandwidth.

In this work, we analyze the Purcell filter of Ref.\
\cite{Jeffrey-14} using both semiclassical and quantum approaches
and considering both the weak and the strong drive regimes. Our
semiclassical analysis uses somewhat different language compared
to the analysis in Ref.\ \cite{Jeffrey-14}; however, the results are
very similar (they show that with the filter the Purcell rate can be
suppressed by two orders of magnitude, while maintaining the same
measurement time). The results of the quantum analysis in the regime
of a weak measurement drive or no drive (considering the
single-photon subspace) practically coincide with the semiclassical
results. In the presence of strong microwave drive, the Purcell rate
is further suppressed with increasing drive strength. We have found
that this suppression is stronger than that obtained without a
filter \cite{Set14}.

In Sec.\ II we discuss the general idea of the bandpass Purcell
filter and analyze its operation semiclassically. Section III is
devoted to the quantum calculation of the Purcell rate in the
presence of the bandpass Purcell filter. In Sec.\ IV we discuss further
suppression of the Purcell rate due to an applied microwave drive.
Section V is the Conclusion. In the appendix we review the basic theory of a
transmon/Xmon qubit measurement, the Purcell decay, and the
corresponding measurement error without the Purcell filter.

\section{Idea of a bandpass Purcell filter and semiclassical analysis}

\begin{figure}
\includegraphics[width=8cm]{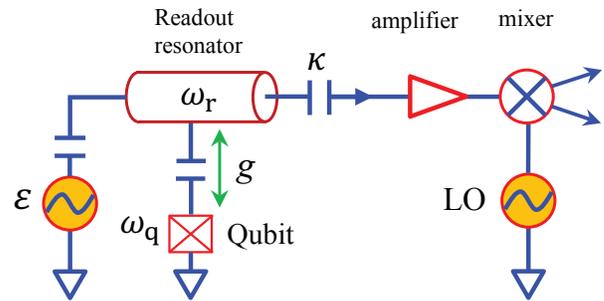}
\caption{Schematic of a standard circuit QED qubit readout setup. The qubit state slightly changes the resonator frequency $\omega_{\rm r}$ (due to qubit-resonator interaction with strength $g$), and this is sensed by passing the microwave through (or reflecting from) the resonator. The amplified outgoing microwave is combined with the local oscillator at the mixer, whose output is measured to discriminate the qubit states. The energy decay $\kappa$ of the resonator is mainly due to its coupling with the transmission line.
  }\label{fig1}
\end{figure}

In the standard cQED setup of dispersive measurement (Fig.\
\ref{fig1}) the qubit interaction with the resonator slightly
changes the effective resonator frequency depending on the qubit
state, so that it is $\omega_{\rm r}^{|e\rangle}$ when the qubit is
in the excited state and $\omega_{\rm r}^{|g\rangle}$ when the qubit
is in the ground state. The dispersive coupling $\chi$ is defined as
    \be
   \chi \equiv (\omega_{\rm r}^{|e\rangle}-\omega_{\rm r}^{|g\rangle})/2 .
    \label{chi-def-text}\ee
In the two-level approximation for the qubit, $\chi=g^2/(\omega_{\rm
q}^{\rm b}-\omega_{\rm r}^{\rm b})$, where $g$ is the
qubit-resonator coupling and $\omega_{\rm q}^{\rm b}$ and
$\omega_{\rm r}^{\rm b}$ are the bare frequencies of the qubit and
the resonator, respectively \cite{Blais-04}. For a transmon or an Xmon qubit,
$\chi$ is usually significantly smaller, $\chi \approx -g^2
\delta_{\rm q}/[(\omega_{\rm q}^{\rm b}-\omega_{\rm r}^{\rm
b})(\omega_{\rm q}^{\rm b}-\delta_{\rm q}-\omega_{\rm r}^{\rm b})]$,
where $\delta_{\rm q}$ is the qubit anharmonicity ($\delta_{\rm q}>0$); moreover, $\chi$
as well as the central frequency $(\omega_{\rm
r}^{|e\rangle}+\omega_{\rm r}^{|g\rangle})/2$ depend on the number
of photons $n$ in the resonator (see \cite{Koch-07,Boisson-10} and the Appendix for a more detailed discussion). The resonator frequency change (and thus the qubit state) is sensed by applying the microwave field with amplitude $\varepsilon$, then amplifying the transmitted or reflected signal, and then mixing it with the applied microwave field to measure its phase and amplitude (Fig.\ \ref{fig1}).

    In the process of measurement, the qubit decays with the Purcell
rate \cite{Blais-04}
    \be
    \Gamma \approx \kappa \,
    \frac{g^2}{(\omega_{\rm q}-\omega_{\rm r})^2},
    \label{Gamma-text}\ee
where $\kappa$ is the resonator energy damping rate (mostly due to leakage into the transmission line -- see Fig.\ \ref{fig1}). Note that in this formula we do
not distinguish the bare and effective frequencies. In the quantum
language this can be interpreted as the leakage with the rate
$\kappa$ of the qubit ``tail'' $g^2/(\omega_{\rm q}-\omega_{\rm
r})^2$, existing in the form of the resonator photon. However, the
Purcell decay also has a simple classical interpretation via the resistive damping
\cite{Esteve-86}, essentially being a linear effect, in contrast to
the dispersion (\ref{chi-def-text}) -- see Appendix for more
details, including dependence of $\Gamma$ on $n$ \cite{Set14}.

The Purcell decay leads to measurement error; therefore, it is
important to reduce the rate $\Gamma$. This can be done by
decreasing the ratio $g/|\omega_{\rm q}-\omega_{\rm r}|$; however,
this decreases $\chi$ and thus increases the necessary measurement
time $t_{\rm m}$ (see Appendix for more details). Another way to
decrease $\Gamma$ is to use a very small leakage rate $\kappa$;
however, this also increases the measurement time $t_{\rm m}$
because the ring-up and ring-down processes give a natural
limitation $t_{\rm m} \agt 4\,\kappa^{-1}$, and in many practical
cases it is even $t_{\rm m} \gg 10 \,\kappa^{-1}$.

It would be good if the rate $\kappa$ which governs the measurement time were
different from $\kappa$ in Eq.\ (\ref{Gamma-text}): specifically if
$\kappa$ for the Purcell decay were much smaller than $\kappa$ for the measurement. This
is exactly what is achieved by using the bandpass filter of Ref.\
\cite{Jeffrey-14}. There are other ways to explain how this Purcell
filter works \cite{Jeffrey-14}, but here we interpret the main idea
of the bandpass Purcell filter as producing different effective
rates $\kappa_{\rm eff}$ for the measurement and for the Purcell decay (so
that the measurement microwave easily passes through the filter,
while the propagation of the photon emitted by the qubit is strongly
impeded by the filter).

\begin{figure}
\includegraphics[width=8cm]{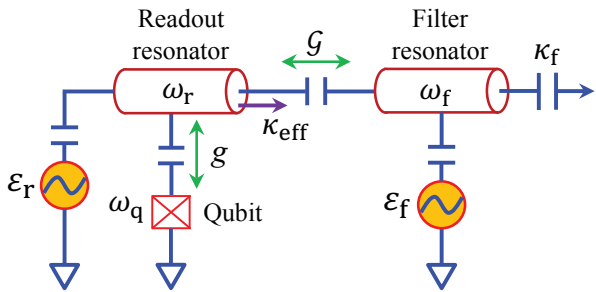}
\caption{Qubit measurement schematic with the bandpass Purcell
filter  of Ref.\ \cite{Jeffrey-14}. The readout resonator with
frequency $\omega_{\rm r}$ (which depends on the qubit state) is
coupled (coupling $\mathcal{G}$) with a filter resonator of
frequency $\omega_{\rm f}$, which decays into the transmission line
with the rate $\kappa_{\rm f}$. The further processing of the
outgoing microwave (not shown) is the same as in Fig.\ \ref{fig1}.
The microwave drive can be applied either to the readout
($\varepsilon_{\rm r}$) or to the filter ($\varepsilon_{\rm f}$)
resonators. Coupling with the decaying filter resonator produces an
effective decay rate $\kappa_{\rm eff}$ of the readout resonator,
which depends on the drive frequency. As a result, for the
measurement microwave $\kappa_{\rm eff}=\kappa_{\rm r}$, while the
qubit sees a much smaller value $\kappa_{\rm eff}=\kappa_{\rm q}$,
thus leading to a suppression of the qubit Purcell decay by a factor $\kappa_{\rm q}/\kappa_{\rm r}$.
  }\label{fig2}
\end{figure}

The schematic of the qubit measurement with the bandpass Purcell
filter of Ref.\ \cite{Jeffrey-14} is shown in Fig.\ \ref{fig2}.
Besides the readout resonator with qubit-state-dependent frequency $\omega_{\rm
r}=\omega_{\rm r}^{|e\rangle}$ or $\omega_{\rm r}=\omega_{\rm
r}^{|g\rangle}$, there is a second (filter)
resonator with frequency $\omega_{\rm f}$, coupled with the readout
resonator with the coupling $\mathcal{G}$. (The coupling
$\mathcal{G}$ is inductive, but we draw it as capacitive to keep the figure simple.) The filter resonator leaks the microwave into the
transmission line with a relatively large damping rate $\kappa_{\rm f}$, so
that its $Q$-factor is $Q_{\rm f}=\omega_{\rm f}/\kappa_{\rm f}
\simeq 30$, while $|\mathcal{G}|\ll \kappa_{\rm f}$. The leaked
field is then amplified and sent to the mixer (not shown) in the
same way as in the standard cQED setup. The readout and filter
resonators are in general detuned from each other, but not much,
$|\omega_{\rm r}-\omega_{\rm f}|\alt \kappa_{\rm f}$ (detuning is needed to
multiplex readout of several qubits using the same filter resonator
\cite{Jeffrey-14,Kelly-15}; for simplicity we consider the measurement of only one qubit). The filter resonator is pumped with the
drive frequency $\omega_{\rm d}$ (close to $\omega_{\rm r}$) and
amplitude $\varepsilon_{\rm f}$. However, for us it will be easier
to first assume instead that the readout resonator is pumped with
amplitude $\varepsilon_{\rm r}$ (Fig.\ \ref{fig2}), and then show
the correspondence between the drives $\varepsilon_{\rm r}$ and
$\varepsilon_{\rm f}$.

Let us use the rotating wave approximation \cite{Allen-Eberly-book,RWA-note} with the rotating frame $e^{-i\omega_d t}$ based on the drive
frequency $\omega_{\rm d}$. Then the evolution of the {\it classical}
field amplitudes $\alpha (t)$ and $\beta (t)$ in the readout and
filter resonators, respectively, is given by the equations
    \begin{align}
& \dot {\alpha} =-i\Delta_{\rm rd} \alpha -i\mathcal{G}\beta
-i\varepsilon_{\rm r} ,
    \label{alpha-dot-1} \\
&  \dot {\beta} = -i\Delta_{\rm fd}\beta -i\mathcal{G}^*\alpha
-\frac{\kappa_{\rm f}}{2}\, \beta , \label{beta-dot-1}
    \end{align}
where $\alpha$ and $\beta$ are normalized so that $|\alpha|^2$ and
$|\beta|^2$ are the average number of photons in the resonators,
$\varepsilon_{\rm r}$ is normalized correspondingly, and
    \be
    \Delta_{\rm rd} = \omega_{\rm r}- \omega_{\rm d}, \,\,\,
  \Delta_{\rm fd} = \omega_{\rm f}- \omega_{\rm d}
    \ee
(recall that $\omega_{\rm r}$ depends on the qubit state). If we are
not interested in the details of evolution on the fast time scale
$\kappa_{\rm f}^{-1}$, then we can use the quasisteady state for
$\beta$ [obtained from Eq.\ (\ref{beta-dot-1}) using $\dot \beta=0$],
    \be
    \beta = \frac{-i \mathcal{G}^*}
    {\kappa_{\rm f}/2 + i \Delta_{\rm fd}}\, \alpha,
    \label{beta-qst}\ee
which can then be inserted into Eq.\ (\ref{alpha-dot-1}), giving
    \begin{eqnarray}
&& \dot {\alpha} = -i(\Delta_{\rm rd}+\delta \omega_{\rm r}) \,
\alpha -\frac{\kappa_{\rm eff}}{2} \, \alpha -i\varepsilon_{\rm r},
    \label{alpha-dot-2}\\
&& \kappa_{\rm eff} = \frac{4 |\mathcal{G}|^2}{\kappa_{\rm f}} \,
\frac{1}{1+(2\Delta_{\rm fd}/\kappa_{\rm f})^2} ,
    \label{kappa-eff}\\
&& \delta\omega_{\rm r} = - \frac{ |\mathcal{G}|^2  \Delta_{\rm
fd}}{(\kappa_{\rm f} /2)^2 +\Delta_{\rm fd}^2}=-\frac{\Delta_{\rm
fd}}{\kappa_{\rm f}}\, \kappa_{\rm eff} .
    \label{delta-omega-r}
    \end{eqnarray}
Thus we see that the field $\alpha$ evolves in practically the same
way as in the standard setup of Fig.\ \ref{fig1}; however,
interaction with the filter resonator shifts the readout resonator
frequency by $\delta\omega_{\rm r}$  and introduces the effective
leakage rate $\kappa_{\rm eff}$ of the readout resonator.

Most importantly, $\kappa_{\rm eff}$ depends on the drive frequency.
For measurement we use $\omega_{\rm d} \approx \omega_{\rm r}$, so
$\kappa_{\rm eff}$ is approximately
    \be
\kappa_{\rm r} \equiv \frac{4 |\mathcal{G}|^2}{\kappa_{\rm f}}\,
\frac{1}{1+[2(\omega_{\rm r}-\omega_{\rm f})/\kappa_{\rm f}]^2} .
    \label{kappa-r}\ee
However, when the qubit tries to leak its excitation through the
readout resonator, this can be considered as a drive at the qubit
frequency, $\omega_{\rm d}=\omega_{\rm q}$, and the corresponding
$\kappa_{\rm eff}$ is then
    \be
\kappa_{\rm q} \equiv \frac{4 |\mathcal{G}|^2}{\kappa_{\rm f}}\,
\frac{1}{1+[2(\omega_{\rm q}-\omega_{\rm f})/\kappa_{\rm f}]^2},
    \label{kappa-q}\ee
which is much smaller than $\kappa_{\rm r}$ if the qubit is detuned
away from the filter linewidth, $|\omega_{\rm q}-\omega_{\rm f}| \gg
\kappa_{\rm f}$. This difference is exactly what we wished for  suppressing the
Purcell rate $\Gamma$: the measurement is governed by $\kappa_{\rm
r}$, while the qubit sees a much smaller value $\kappa_{\rm q}$.
Therefore, we would expect that the Purcell rate is given by Eq.\
(\ref{Gamma-text}) with $\kappa=\kappa_{\rm q}$ [see Eq.\ (\ref{Gamma-wf}) later], while the
``separation'' measurement error is given by Eqs.\
(\ref{delta-alpha})--(\ref{P-err-sep}) with $\kappa=\kappa_{\rm r}$
(see Appendix). As a result, compared with the standard setup (Fig.\
\ref{fig1}) with the same physical parameters for measurement, the
Purcell rate is suppressed by the factor
    \be
    F= \frac{\kappa_{\rm q}}{\kappa_{\rm r}} =
    \frac{1+[2(\omega_{\rm r}-\omega_{\rm f})/\kappa_{\rm f}]^2}{1+[2(\omega_{\rm q}-\omega_{\rm f})/\kappa_{\rm f}]^2} \ll 1.
        \label{F-1}\ee
This is essentially the {\it main result} of this paper, which will
be confirmed by the quantum analysis in the next section. (To avoid a possible confusion, we note that $\kappa_{\rm q}$ is not the qubit decay rate; it is the resonator decay, as seen by the qubit.)

Our result for the Purcell suppression factor was based on the
behavior of the field amplitude in the readout resonator. Let us
also check that the field $\gamma_{\rm tl}$ propagating in the outgoing
transmission line behaves according to the effective model as well.
The outgoing field amplitude is $\gamma_{\rm tl}=\sqrt{\kappa_{\rm
f}}\, \beta$ (in the normalization for which $|\gamma_{\rm tl}|^2$ is
the average number of propagating photons per second). Using Eq.\
(\ref{beta-qst}), we find
    \be
    \gamma_{\rm tl} = \frac{-i \mathcal{G}^* \sqrt{\kappa_{\rm f}}}
    {\kappa_{\rm f}/2 + i \Delta_{\rm fd}}\, \alpha = e^{i \varphi}
    \sqrt{\kappa_{\rm eff}} \, \alpha,
    \label{gamma-tl-1}\ee
so, as expected, the outgoing amplitude behaves as in the standard
setup of Fig.\ \ref{fig1} with $\kappa=\kappa_{\rm eff}$, up to an
unimportant phase shift $\varphi= {\rm arg} [-i
\mathcal{G}^*/(\kappa_{\rm f}/2 + i \Delta_{\rm fd})]$. Note that to
show the equivalence between the dynamics (including transients) of
the systems in Figs.\ \ref{fig1} and \ref{fig2} we needed the
assumption of a sufficiently large $\kappa_{\rm f}$ in order to use
the quasisteady state (\ref{beta-qst}). However, this assumption is
not needed if we consider only the steady state (without
transients).

So far we assumed that the measurement is performed by driving the
readout resonator with the amplitude $\varepsilon_{\rm r}$. Now let
us consider the realistic case \cite{Jeffrey-14, Kelly-15} when the
drive $\varepsilon_{\rm f}$ is applied to the filter resonator. The evolution equations (\ref{alpha-dot-1}) and (\ref{beta-dot-1}) for the classical field amplitudes are then replaced by
    \begin{align}
& \dot {\alpha} =-i\Delta_{\rm rd} \alpha -i\mathcal{G}\beta ,
    \label{alpha-dot-3} \\
&  \dot {\beta} = -i\Delta_{\rm fd}\beta -i\mathcal{G}^*\alpha
-\frac{\kappa_{\rm f}}{2}\, \beta -i\varepsilon_{\rm f} ,
\label{beta-dot-3}
    \end{align}
so that the quasisteady state for the filter resonator is
    \be
    \beta = \frac{-i \mathcal{G}^* }
    {\kappa_{\rm f}/2 + i \Delta_{\rm fd}}\, \alpha  + \frac{ -i\varepsilon_{\rm f}}
    {\kappa_{\rm f}/2 + i \Delta_{\rm fd}} ,
    \label{beta-qst-2}\ee
and the field evolution in the readout resonator is
    \be
     \dot {\alpha} = -i(\Delta_{\rm rd}+\delta \omega_{\rm r}) \,
\alpha -\frac{\kappa_{\rm eff}}{2} \, \alpha -
\frac{\mathcal{G}}{\kappa_{\rm f}/2 + i \Delta_{\rm fd}} \,
\varepsilon_{\rm f},
    \label{alpha-dot-4}
    \ee
with the same $\kappa_{\rm eff}$ and $\delta\omega_{\rm r}$  given
by Eqs.\ (\ref{kappa-eff}) and (\ref{delta-omega-r}). The only
difference between the effective evolution equations
(\ref{alpha-dot-2}) and (\ref{alpha-dot-4}) is a linear relation,
    \be
\varepsilon_{\rm r} \leftrightarrow -i  \varepsilon_{\rm f}
\mathcal{G} /(\kappa_{\rm f}/2 + i \Delta_{\rm fd}),
    \label{er-ef}\ee
 between the
drive amplitudes $\varepsilon_{\rm r}$ and $\varepsilon_{\rm f}$
producing the same effect. Therefore, our results obtained above remain
unchanged for driving the filter resonator, and the Purcell rate
suppression factor is still given by Eq.\ (\ref{F-1}).

Note that in the quasisteady state the separation between the filter
amplitudes $\beta$ for the two qubit states does not depend on
whether the drive is applied to the filter or readout  resonator, as
long as we use the correspondence (\ref{er-ef}) between the drive
amplitudes. The same is true for the separation between the outgoing
fields $\gamma_{\rm tl}$. Similarly, the separation between the
outgoing fields for the two qubit states is the same (up to  the
phase $\varphi$) as in the standard setup of Fig.\ \ref{fig1} with
$\varepsilon=\varepsilon_{\rm r}$, $\kappa =\kappa_{\rm r}$, and the
resonator frequency adjusted by $\delta\omega_{\rm r}$ given by Eq.\
(\ref{delta-omega-r}). Therefore, these configurations are
equivalent to each other from the point of view of quantum
measurement, including interaction between the qubit and readout
resonator, extraction of quantum information, back-action, etc.

Nevertheless, driving the filter resonator produces a different
outgoing field $\gamma_{\rm tl}=\sqrt{\kappa_{\rm f}}\, \beta$,
which now contains an additional term $-i\varepsilon_{\rm f}
\sqrt{\kappa_{\rm f}} / (\kappa_{\rm f}/2 + i \Delta_{\rm fd})$ in
comparison with Eq.\ (\ref{gamma-tl-1}), which comes from the second
term in Eq.\ (\ref{beta-qst-2}). In particular, instead of the
Lorentzian line shape of the transfer function when driving the
readout resonator, the transfer function for driving the filter is
(in the steady state)
   \be
   \frac{\gamma_{\rm tl}^{\rm (f)}}{\varepsilon_{\rm f}}=
  \frac{ \sqrt{\kappa_{\rm f}}}{\kappa_{\rm f}/2+i\Delta_{\rm fd}}
  \, \frac{ 2\Delta_{\rm rd}/\kappa_{\rm eff}}
  {\displaystyle 1 +\frac{2i (\Delta_{\rm rd}+\delta\omega_{\rm r})}{\kappa_{\rm eff}}},
    \label{line-shape}\ee
where $\kappa_{\rm eff}$ can be replaced with $\kappa_{\rm r}$. (Note a non-standard normalization of the transfer function because of different normalizations of $\gamma_{\rm tl}^{\rm (f)}$ and $\varepsilon_{\rm f}$.) This
line shape for the amplitude $|\gamma_{\rm tl}^{\rm (f)}/\varepsilon_{\rm f}|$ shows a dip near $\omega_{\rm r}$ (note that $\gamma_{\rm
tl}^{\rm (f)}/\varepsilon_{\rm f}=0$ at $\omega_{\rm d}=\omega_{\rm
r}$) and is  significantly asymmetric when $\delta\omega_{\rm r}$ is
comparable to $\kappa_{\rm r}$; this occurs when the detuning
between the readout and filter resonators is comparable to
$\kappa_{\rm f}$ -- see Eq.\ (\ref{delta-omega-r}). In terms of the
field $\alpha$ in the readout resonator, the outgoing field at
steady state is
    \be
    \gamma_{\rm tl}^{\rm (f)} = - \frac{\sqrt{\kappa_{\rm f}} \,
    \Delta_{\rm rd}}{\mathcal{G}} \, \alpha
\,\,\,
    \label{gamma-tl-2}\ee
instead of Eq.\ (\ref{gamma-tl-1}) for driving the readout resonator.

The difference between the outgoing fields  $\gamma_{\rm tl}^{\rm (f)}$ and  $\gamma_{\rm tl}^{\rm (r)}$ when driving the filter or readout resonator (for the same $\alpha$, i.e., the same measurement conditions) may be important for saturation of the microwave amplifier. The ratio of the corresponding outgoing
powers is
    \be
 \frac{| \gamma_{\rm tl}^{\rm (f)}|^2}{| \gamma_{\rm tl}^{\rm
 (r)}|^2} = \left(\frac{\Delta_{\rm rd}}
 {\kappa_{\rm r}}\right)^2 \frac{4}{1+(2\Delta_{\rm fd} /\kappa_{\rm f})^2}  ,
 \,\,\,\,
    \label{power-ratio}\ee
where we assumed $|\Delta_{\rm rd}|\ll
\kappa_{\rm f}$ (so that $\kappa_{\rm eff}\approx \kappa_{\rm r}$).
For example, if the drive frequency is chosen as $\omega_{\rm
d}=(\omega_{\rm r}^{|g\rangle}+\omega_{\rm r}^{|e\rangle})/2$, then
$\Delta_{\rm rd}=\pm \chi$; if in this case $|\chi| \ll \kappa_{\rm
r}$, then driving the filter resonator is advantageous because it
produces less power to be amplified, while driving the readout
resonator is advantageous if $|\chi| \gg \kappa_{\rm r}$. However,
when $\omega_{\rm d} \neq (\omega_{\rm r}^{|g\rangle}+\omega_{\rm
r}^{|e\rangle})/2$, the situation is more complicated.

\begin{figure}
\includegraphics[width=8cm]{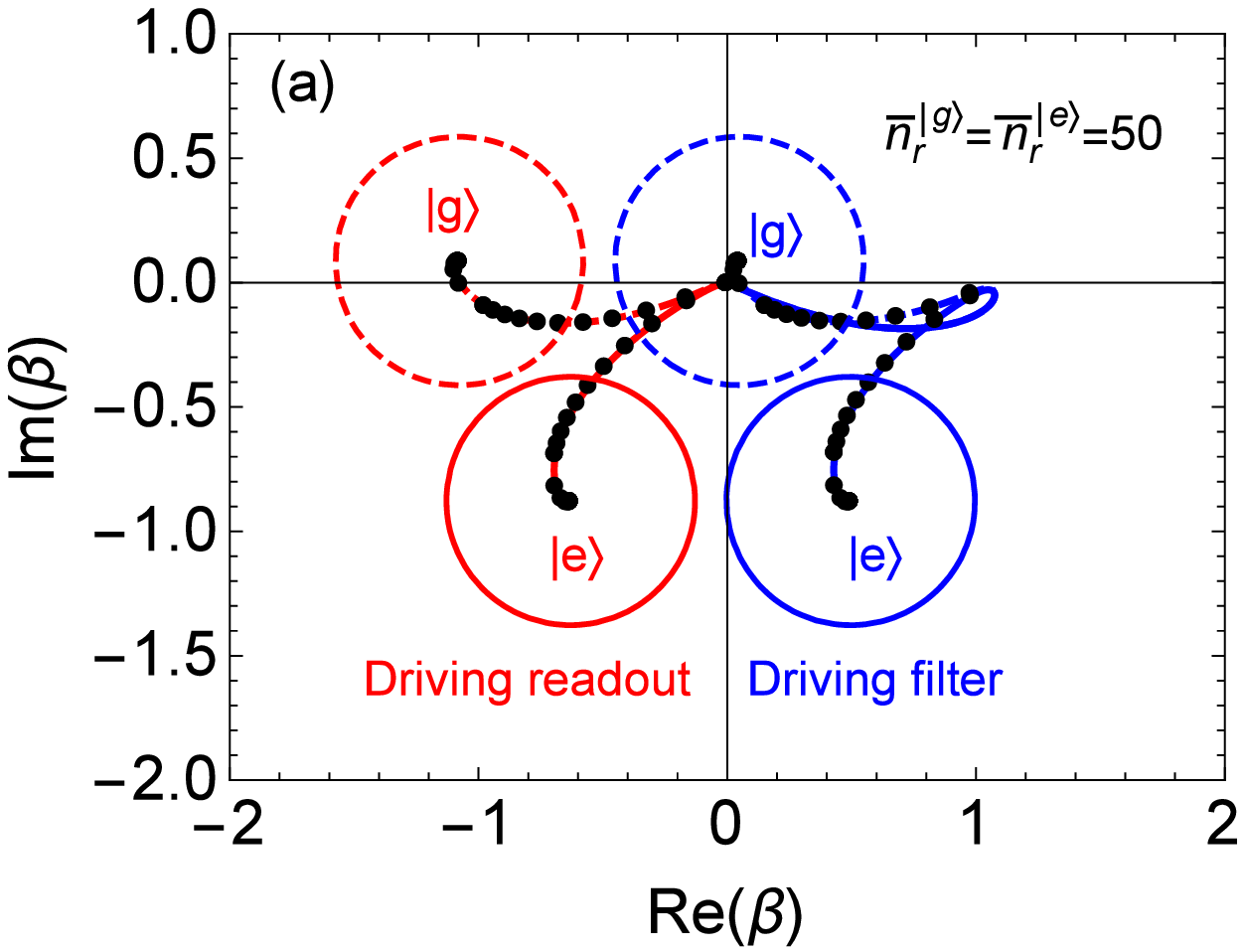}
\includegraphics[width=8cm]{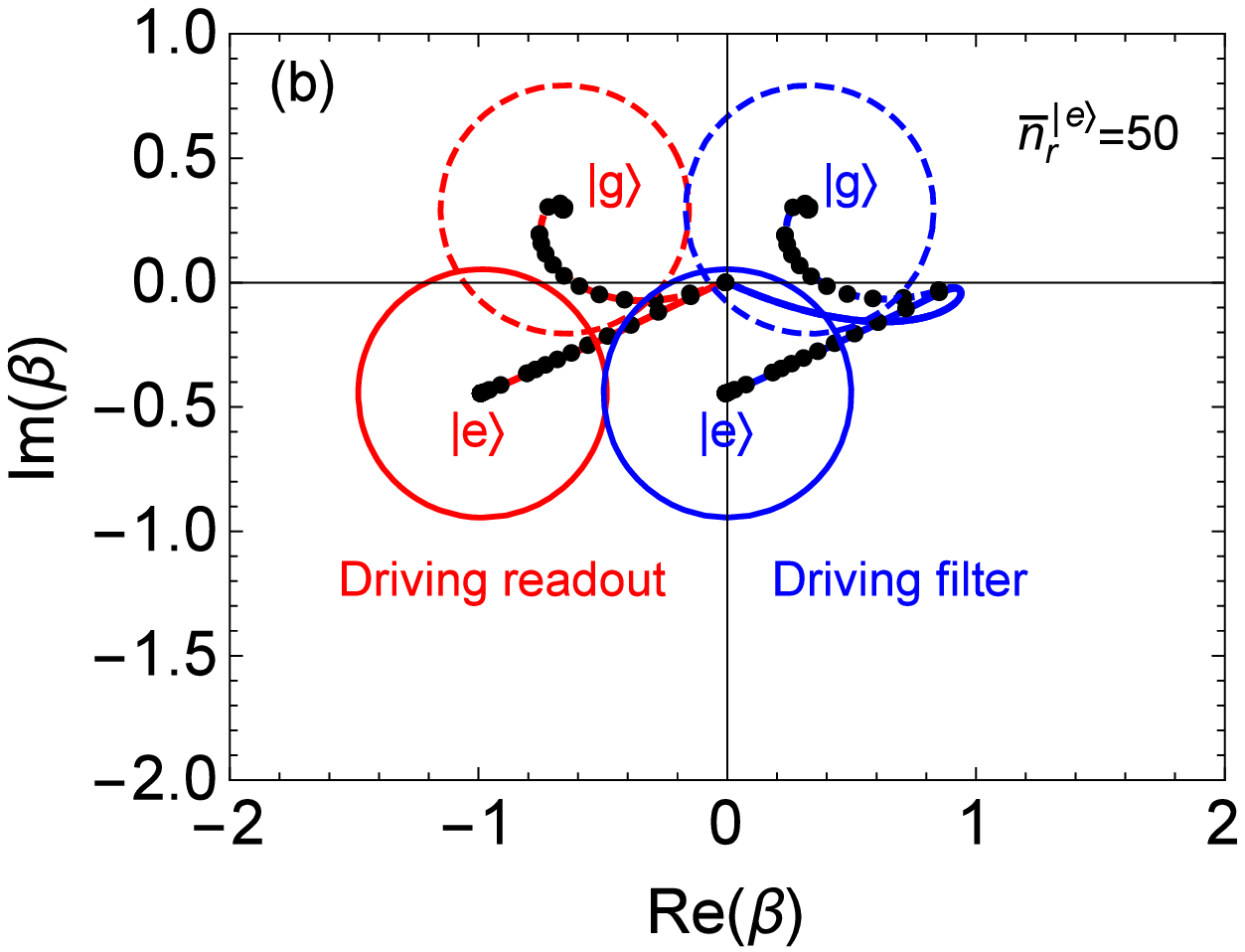}
\caption{Phase-space transient evolution of qubit-state-dependent
coherent states in the filter resonator for driving either the
readout resonator (red curves, left) or the filter (blue curves, right), with the
drive amplitudes related via Eq.\ (\ref{er-ef}). Solid curves are
for the qubit state $|e\rangle$, dashed curves are for the state
$|g\rangle$. See text for the assumed parameters. For panel (a) we
choose the drive frequency $\omega_{\rm d}$ symmetrically for the
readout resonator, so that  $\bar{n}_{\rm
r}^{|g\rangle}=\bar{n}_{\rm r}^{|e\rangle}=50$. For panel (b) we
choose $\omega_{\rm d}$ symmetrically for the filter resonator, so that
$\bar{n}_{\rm f}^{|g\rangle}=\bar{n}_{\rm f}^{|e\rangle}$ when
driving the filter; we choose $\bar{n}_{\rm r}^{|e\rangle}=50$. The
black dots indicate time moments in the evolution every 10 ns until 100 ns, then
every 50 ns. Circles illustrate the coherent state error circles in the steady state.
} \label{fig3}
\end{figure}

Figure \ref{fig3} shows the transient phase-space evolution $\beta
(t)$ of the field (coherent state) in the filter resonator when a
step-function drive is applied to the readout resonator (red curves)
or to the filter resonator (blue curves), with the drive amplitudes
related via Eq.\ (\ref{er-ef}), and for real $\varepsilon_{\rm r}$. The
solid curves show the evolution when the qubit is in the excited
state, while the dashed curves are for the qubit in the ground
state. We have used parameters similar to the experimental
parameters of Ref.\ \cite{Jeffrey-14}: $\omega_{\rm
r}^{|e\rangle}/2\pi=6.8$ GHz, $\omega_{\rm
r}^{|g\rangle}/2\pi=6.803$ GHz (so that $2\chi/2\pi=-3$ MHz),
$\omega_{\rm f}/2\pi = 6.75$ GHz, $Q_{\rm f}=30$ (so that
$\kappa_{\rm f}^{-1}= 0.71$ ns), and $\kappa_{\rm r}^{-1}=30$ ns (so
that $\mathcal{G}/2\pi =18.9$ MHz). The field evolution is
calculated using either Eqs.\
(\ref{alpha-dot-1})--(\ref{beta-dot-1}) or Eqs.\
(\ref{alpha-dot-3})--(\ref{beta-dot-3}); in simulations we have
neglected the dependence of $\chi$ and $(\omega_{\rm
r}^{|e\rangle}+\omega_{\rm r}^{|g\rangle})/2$ on the average number
of photons $\bar n_{\rm r}=|\alpha |^2$ in the readout resonator (see Appendix).
The black dots indicate the time moments every 10 ns between 0 and
100 ns, and then every 50 ns. The circles illustrate the coherent
state error circles \cite{Scully-book} in the steady state. We see that when the drive is applied to
the filter resonator (blue curves), the evolution $\beta (t)$ is
initially very fast (governed by $\kappa_{\rm f}$), while after the
quasisteady state is reached, the evolution is governed by a much
slower $\kappa_{\rm r}$, eventually approaching the steady state.
When the drive is applied to the readout resonator (red curves), the
transient evolution is always governed by the slower decay
$\kappa_{\rm r}$.

In Fig.\ \ref{fig3}(a) we choose the drive frequency $\omega_{\rm
d}$ symmetrically from the point of view of the readout resonator
field, so that in the steady state $\bar{n}_{\rm
r}^{|g\rangle}=\bar{n}_{\rm r}^{|e\rangle}=50$, where $\bar{n}_{\rm
r}^{|g\rangle}$ and $\bar{n}_{\rm r}^{|e\rangle}$ are the average
photon numbers for the two qubit states. For that we need
$\omega_{\rm d}=(\omega_{\rm r}^{|g\rangle}+\omega_{\rm
r}^{|e\rangle})/2+\delta\omega_{\rm r}$ with $\delta\omega_{\rm r}$
given by Eq.\ (\ref{delta-omega-r}); for our parameters
$\delta\omega_{\rm r}/2\pi =1.23$ MHz, so $\omega_{\rm d}/2\pi =
6.80273$ GHz. Such symmetric choice of the drive frequency provides
the largest separation between the two coherent states for a fixed
drive amplitude if $2 \,|\chi|<\kappa_{\rm r}$. While the field in
the readout resonator is always symmetric in this case, Fig.\
\ref{fig3}(a) shows that the field in the filter resonator is
symmetric only when driving the readout resonator (red curves,
$\bar{n}_{\rm f}^{|g\rangle}=\bar{n}_{\rm f}^{|e\rangle}=1.2$),
while it is strongly asymmetric when the filter resonator is driven (blue
curves, $\bar{n}_{\rm f}^{|g\rangle}=0.01$, $\bar{n}_{\rm
f}^{|e\rangle}=1.0$; $\bar{n}_{\rm f}^{|g\rangle}$ is very small
because for our parameters $\delta\omega_{\rm r}\approx |\chi|$ and therefore $\omega_{\rm d}\approx \omega_{\rm
r}^{|g\rangle}$). The number of photons in the filter is much less
than in the readout resonator because $\kappa_{\rm f} \gg
\kappa_{\rm r}$.

In Fig.\ \ref{fig3}(b) we choose $\omega_{\rm d}$ so that in the
steady state $\bar{n}_{\rm f}^{|g\rangle}=\bar{n}_{\rm
f}^{|e\rangle}$ for driving the filter; this is the natural choice
for decreasing the microwave power to be amplified. This occurs at
$\omega_{\rm d}/2\pi =6.80120$, which is close to the expected value
$(\omega_{\rm r}^{|g\rangle}+\omega_{\rm r}^{|e\rangle})/2$, but not
equal because of the asymmetry of the line shape (\ref{line-shape}).
We choose the amplitudes to produce $\bar{n}_{\rm r}^{|e\rangle}=50$
(then $\bar{n}_{\rm r}^{|g\rangle}=22$). The difference between
$\bar{n}_{\rm r}^{|e\rangle}$ and $\bar{n}_{\rm r}^{|g\rangle}$
leads to different values $\bar{n}_{\rm f}^{|e\rangle}=1.2$ and
$\bar{n}_{\rm f}^{|g\rangle}=0.5$ when driving the readout
resonator, while for driving the filter the field in the filter is
symmetric, $\bar{n}_{\rm f}^{|e\rangle}=\bar{n}_{\rm
f}^{|g\rangle}=0.2$. Compared to the case of Fig.\  \ref{fig3}(a),
there is 5 times less power to be amplified for the $|e\rangle$
state (when driving the filter); however, the state separation is
1.3  times smaller (in amplitude) for the same $\bar{n}_{\rm
r}^{|e\rangle}$. Thus, there is a trade-off between the state
separation and amplified power in choosing the drive frequency.
Comparing the red and blue curves in Fig.\ \ref{fig3}(b), we see that in the steady state $\bar{n}_{\rm f}^{|g\rangle}$ and  $\bar{n}_{\rm f}^{|e\rangle}$  are smaller for driving the
filter rather than the readout resonator. This is beneficial because
there is less power to be amplified; however, the ratio is not very big (as
expected for a moderate value $|\chi|/\kappa_{\rm r}=0.28$).

Note that our definition of $\kappa_{\rm r}$ in Eq.\
(\ref{kappa-r}) is not strictly well-defined because the resonator
frequency $\omega_{\rm r}$ depends on the qubit state, and the
drive frequency can be in between $\omega_{\rm r}^{\rm |e\rangle}$
and $\omega_{\rm r}^{\rm |g\rangle}$. However, this frequency
difference is much smaller than $\kappa_{\rm f}$, and therefore not
important for practical purposes in the definition of $\kappa_{\rm
r}$. In an experiment $\kappa_{\rm r}$ can be measured either via the field decay
\cite{Jeffrey-14} or via the linewidth of the steady-state transfer
function showing the dip of $|\gamma_{\rm tl}^{\rm (f)}/\varepsilon_{\rm f}|$ near the resonance $\omega_{\rm d}=\omega_{\rm r}$  [Eq.\ (\ref{line-shape})], since near the dip $\kappa_{\rm eff}\approx \kappa_{\rm r}$.

Thus far we assumed that all decay $\kappa_{\rm f}$ in the filter
resonator is due to the leakage $\kappa_{\rm f}^{\rm out}$ into the
outgoing transmission line. If $\kappa_{\rm f}^{\rm out} <
\kappa_{\rm f}$ and the decay $\kappa_{\rm f}-\kappa_{\rm f}^{\rm
out}$ is due to leakage into the line delivering the drive
$\varepsilon_{\rm f}$ or due to another dissipation channel, then
the only difference compared to the previous discussion is the extra
factor $\sqrt{\kappa_{\rm f}^{\rm out}/\kappa_{\rm f}}$ for the
outgoing field $\gamma_{\rm tl}$. This will lead to multiplication
of the overall quantum efficiency of the measurement by $\kappa_{\rm
f}^{\rm out}/\kappa_{\rm f}$ and will only slightly affect the
measurement fidelity.  Adding dissipation in the readout resonator with
rate $\kappa_{\rm r,d}$ increases the effective linewidth to
$\kappa_{\rm eff}+\kappa_{\rm r,d}$ and multiplies the quantum
efficiency by $\kappa_{\rm r}/(\kappa_{\rm r}+\kappa_{\rm r,d})$.
Most importantly, since $\kappa_{\rm r,d}$ does not change with
frequency, the Purcell suppression factor (\ref{F-1}) becomes
$(\kappa_{\rm q}+\kappa_{\rm r,d})/(\kappa_{\rm r}+\kappa_{\rm
r,d})$, so that the Purcell filter performance deteriorates; we will
discuss this in a little more detail in Sec.\ III$\,$C.

Note that our main result (\ref{F-1}) for the Purcell suppression
factor is slightly different from the result $F=[\kappa_{\rm
f}/2(\omega_{\rm q}-\omega_{\rm r})]^2 (\omega_{\rm q}/\omega_{\rm
r})^2$, which was derived in Ref.\ \cite{Jeffrey-14} using the
circuit theory. The reasons are the following. First, in the
derivation of \cite{Jeffrey-14} it was assumed that the two
resonators have the same frequency, which makes the numerator in
Eq.\ (\ref{F-1}) equal to 1. Second, the term 1 in the denominator
in Eq.\ (\ref{F-1}) was essentially neglected in comparison with the
larger second term. Finally, the role of the factor $(\omega_{\rm
q}/\omega_{\rm r})^2$ is not quite clear. In the derivation of Ref.\
\cite{Jeffrey-14} keeping this factor was exceeding the accuracy of
the derivation, while in our derivation we essentially use the
rotating wave approximation, which assumes $\omega_{\rm
q}/\omega_{\rm r}\approx 1$. Aside from these small differences, our
result coincides with the result of Ref.\ \cite{Jeffrey-14}.


\section{Quantum analysis in single-excitation subspace}

In this section we discuss the quantum derivation of the Purcell rate in the presence of the
bandpass filter in the regime when the resonators are not driven or
driven sufficiently weakly to neglect dependence of the Purcell rate
on the number of photons in the resonator \cite{Set14}. More
precisely, we consider the quantum evolution in the
single-excitation (and zero-excitation) subspace. We apply two
methods: the wavefunction approach, in which we use a non-Hermitian
Hamiltonian with a decaying wavefunction, and the more traditional
density matrix analysis.

In the absence of the drive and in the rotating wave approximation, the relevant Hamiltonian of the system
shown in Fig.\ \ref{fig2} (without considering decay $\kappa_{\rm
f}$) is ($\hbar =1$)
    \begin{eqnarray}
&& H =  \omega_{\rm q}^{\rm b} \sigma_+\sigma_- + \omega_{\rm r}^{\rm b} a^\dagger a + \omega_{\rm f} b^\dagger b + g (a^\dagger \sigma_- + a\sigma_+)
    \nonumber \\
&& \hspace{0.8cm} +\mathcal{G} a^\dagger b + \mathcal{G}^* a b^\dagger ,
    \label{Ham-start}\end{eqnarray}
where $\omega_{\rm q}^{\rm b}$ is the bare qubit frequency, $\omega_{\rm r}^{\rm b}$ is the bare frequency of the readout
resonator, $\omega_{\rm f}$ is the  filter resonator frequency,
raising/lowering operators $\sigma_{+}$ and $\sigma_-$ act on the
qubit state, $a^\dagger$ and $a$ are the creation and annihilation
operators for the readout resonator, $b^\dagger$ and $b$ are for the
filter resonator, $g$ is the qubit-readout resonator coupling, and
$\mathcal{G}$ is the resonator-resonator coupling. For simplicity we
assume a real positive $g$, but $\mathcal{G}$ can be complex for generality (for the capacitive or inductive coupling between the resonators, $\mathcal{G}$ is real if the same generalized coordinates are used for both resonators).

Note that in the case without drive it is sufficient to consider only two
levels for the qubit because only the single-excitation (and
zero-excitation) subspace is involved in the evolution, and therefore the amount of qubit nonlinearity due to the Josephson junction is irrelevant. However, in the presence of a drive (considered in the next section) it is formally necessary to take into account several levels in the qubit (as done in the Appendix). Nevertheless, to leading order the Purcell rate is insensitive to this, because the Purcell decay is essentially a classical linear effect (see discussion in the Appendix). Also note that the lab-frame Hamiltonian (\ref{Ham-start}) assumes the rotating wave approximation (as in the standard Jaynes-Cummings Hamiltonian), since it neglects the ``counter-rotating'' terms of the form $a^\dagger \sigma_+$, $a\sigma_-$, $a^\dagger b^\dagger$, and $ab$. This requires assumption that $|\omega_{\rm q}^{\rm b}-\omega_{\rm r}^{\rm b}|$, $|\omega_{\rm f}-\omega_{\rm r}^{\rm b}|$, $g$, and $|\mathcal{G}|$ are small compared to $\omega_{\rm r}^{\rm b}$.

Let us choose the rotating frame with frequency $\omega_{\rm q}^{\rm b}$, i.e., $H_0=\omega_{\rm q}^{\rm b} (\sigma_+\sigma_- + a^\dagger a+b^\dagger b)$; then the interaction Hamiltonian $V=H-H_0$ is
\begin{equation}\label{h}
  V=\Delta_{\rm rq} a^\dagger a + \Delta_{\rm fq} b^{\dag}b
  +\textit{g}(a^{\dag}\sigma_{-}+a\sigma_{+})+\mathcal{G}a^{\dag}b+ \mathcal{G}^* ab^{\dag},
\end{equation}
where
    \be
\Delta_{\rm rq}=\omega_{\rm r}^{\rm b}-\omega_{\rm q}^{\rm b}, \,\,\,
\Delta_{\rm fq}=\omega_{\rm f}-\omega_{\rm q}^{\rm b},
    \ee
and the interaction picture is equivalent to the Schr\"odinger picture because $\exp(iH_0 t) V\exp(-iH_0 t)=V$, which is because the starting Hamiltonian (\ref{Ham-start}) already assumes the rotating-wave approximation.
 The master
equation for the density matrix $\rho$, which includes the damping $\kappa_{\rm f}$ of the filter resonator is
\begin{equation}\label{ma}
  \dot \rho=-i[V,\rho]+\kappa_{\rm f} \left(b\rho b^{\dag}-b^{\dag}b\rho/2-\rho b^{\dag}b/2\right) .
\end{equation}

In general, the bare basis is $|jnm\rangle$, where $j$ represents the qubit states, while $n$ and $m$ represent the readout and filter resonator Fock states,
respectively. However, in this section we consider only the single-excitation (and zero-excitation) subspace, so only four bare states are relevant: $|\textbf{e}\rangle \equiv |\text{e}00\rangle$, $|\textbf{r}\rangle=|\text{g}10\rangle$, $|\textbf{f}\rangle =|\text{g}01\rangle$, and $|\textbf{g}\rangle=|\text{g}00\rangle$.

Note that the interaction hybridizes the bare states of the qubit and the resonators. (Hybridization of the readout resonator mode is essentially what makes the qubit measurement possible.) Therefore, when discussing the Purcell rate for the qubit energy relaxation, we actually mean decay of the eigenstate, corresponding to the qubit  excited state. This makes perfect sense experimentally, since manipulations of the qubit state usually occur in the eigenbasis (adiabatically, compared with the qubit detuning from the resonator).

\subsection{Method I: Decaying wavefunction}

Instead of using the traditional density matrix language for the
description of the Purcell effect \cite{Haroche-book}, it is easier
to use the language of wavefunctions, even in the presence of the
decay $\kappa_{\rm f}$ \cite{Set14}. Physically, the wave functions
can still be used because in the single-excitation subspace
unraveling of the Lindblad equation corresponds to only one ``no
relaxation'' scenario (see, e.g., \cite{Kor-13}), and therefore the
wavefunction evolution is non-stochastic. Another, more formal way
to introduce this language, is to rewrite the master equation
\eqref{ma} as \cite{Pie07,Car93} $\dot \rho=-i[H_{\rm
eff},\rho]+\kappa_{\rm f} b\rho b^{\dag}$, where $H_{\rm
eff}=V-i\kappa_{\rm f} b^{\dag}b/2$ is an effective non-Hermitian
Hamiltonian. Next, the term $\kappa_{\rm f} b\rho b^{\dag}$ can be
neglected because in the single-excitation subspace it produces only
an ``incoming'' contribution  from higher-excitation subspaces,
which are not populated. Therefore, in the single-excitation
subspace we can use $\dot \rho=-i[H_{\rm eff},\rho]$. Equivalently,
$|\dot \psi \rangle=-iH_{\rm eff}|\psi\rangle$, which describes the
evolution of the {\it decaying} wavefunction
$|\psi(t)\rangle=c_{\textbf e}(t)|\textbf{e}\rangle+c_{\textbf
r}(t)|\textbf{r}\rangle+c_{\textbf f}(t)|\textbf{f}\rangle$.
Therefore, the probability amplitudes
$c_{\textbf{e},\textbf{r},\textbf{f}}$ satisfy the following
equations:
\begin{eqnarray}
  && \dot c_{\textbf e} = -i g c_{\textbf r},
    \label{c-dot-e}\\
  && \dot c_{\textbf r} = -i\Delta_{\rm rq}  c_{\textbf r} -i gc_{\textbf e} -i\mathcal{G} c_{\textbf f},
   \label{c-dot-r}\\
  && \dot c_{\textbf f} = -i \Delta_{\rm fq} c_{\textbf f} -i\mathcal{G}^*c_{\textbf r} -(\kappa_{\rm f}/2)\, c_{\textbf f},
 \label{c-dot-f}\end{eqnarray}
while the population $\rho_{\textbf{gg}}$ of the zero-excitation
state $|\textbf{g}\rangle$ evolves as
$\dot\rho_{\textbf{gg}}=\kappa_{\rm f}|c_{\textbf{ff}}|^2$ or can be
found as $\rho_{\textbf{gg}}=1-|c_{\textbf{ee}}|^2
-|c_{\textbf{rr}}|^2 -|c_{\textbf{ff}}|^2$. Note that Eqs.\
(\ref{c-dot-r}) and (\ref{c-dot-f}) exactly correspond to the
classical equations (\ref{alpha-dot-1}) and (\ref{beta-dot-1}) with
$\Delta_{\rm rd}$ replaced with $\Delta_{\rm rq}$, also $\Delta_{\rm
fd}$ replaced with $\Delta_{\rm fq}$, and $\varepsilon_{\rm r}$
replaced with $gc_{\textbf e}$.

From the eigenvalues $\lambda_{\textbf{e},\textbf{r},\textbf{f}}=-i
\omega_{\textbf{e},\textbf{r},\textbf{f}}-\Gamma_{\textbf{e},\textbf{r},\textbf{f}}/2$
of the matrix representing Eqs.\ (\ref{c-dot-e})--(\ref{c-dot-f}),
one can obtain the  eigenfrequencies
$\omega_{\textbf{e},\textbf{r},\textbf{f}}$ and the corresponding
decay rates $\Gamma_{\textbf{e},\textbf{r},\textbf{f}}$. These
eigenvalues can be found from the qubic equation
\begin{align} \label{lambda-eq}
& \lambda ^3+ \lambda^2 (i\Delta_{\rm rq} +i\Delta_{\rm fq} + \kappa_{\rm f}/2)
+ \lambda (- \Delta_{\rm rq}\Delta_{\rm fq} +|\mathcal{G}|^2 +g^2
    \notag\\
& \hspace{0.5cm} + i \Delta_{\rm rq} \kappa_{\rm f}/2) + g^2 (i \Delta_{\rm fq}+ \kappa_{\rm f}/2 ) =0 .
\end{align}
We are interested in the Purcell rate $\Gamma=\Gamma_{\textbf e}$,
which corresponds to the decay of the eigenstate close to
$|\textbf{e}\rangle$. Since $\lambda_{\textbf{e}}$ is close to zero,
in the first approximation we can neglect the term $\lambda^3$ in
Eq.\ (\ref{lambda-eq}), thus reducing it to the quadratic equation.
If more accuracy is needed, the equation can be solved iteratively,
replacing $\lambda^3$ with the value found in the previous iteration
(the second iteration is usually sufficient).

Besides finding the Purcell rate $\Gamma$ exactly or approximately
from Eq.\ (\ref{lambda-eq}), we can find it approximately by using
quasisteady solutions of Eqs.\ (\ref{c-dot-r}) and (\ref{c-dot-f}),
to a large extent  following the classical derivation in the
previous section. Assuming $\dot c_{\textbf f}=0$ in Eq.\
(\ref{c-dot-f}), we find $c_{\textbf f}=-i\mathcal{G}^* c_{\textbf
r}/(i\Delta_{\rm fq}+\kappa_{\rm f}/2)$. Inserting this quasisteady
value into Eq.\ (\ref{c-dot-r}) and assuming $\dot c_{\textbf r}=0$,
we find $c_{\textbf r}=-igc_{\textbf e}/[i\Delta_{\rm rq} +
|\mathcal{G}|^2 /(i\Delta_{\rm fq}+\kappa_{\rm f}/2)]$. Substituting
this quasisteady value into Eq.\ (\ref{c-dot-e}), we obtain
    \be
    \dot c_{\textbf e}=-\frac{ g^2 }{i\Delta_{\rm rq} + |\mathcal{G}|^2 /(i\Delta_{\rm fq}+\kappa_{\rm f}/2)} \, c_{\textbf e} = \lambda_{\textbf e} c_{\textbf e}.
    \label{lambda-e}\ee
Finally, we obtain the Purcell rate as $\Gamma =-2{\rm
Re}(\lambda_{\textbf e})$,
     \begin{eqnarray}
    && \Gamma = \frac{g^2 |\mathcal{G}|^2\kappa_{\rm f}}
    {\Delta_{\rm rq}^2 [(\Delta_{\rm fq}-|\mathcal G|^2/\Delta_{\rm rq})^2 +(\kappa_{\rm f}/2)^2]}
        \label{Gamma-wf-1}\\
     && \hspace{0.3cm} \approx \frac{g^2 |\mathcal{G}|^2\kappa_{\rm f}}
    {\Delta_{\rm rq}^2 [\Delta_{\rm fq}^2 +(\kappa_{\rm f}/2)^2]} = \frac{g^2\kappa_{\rm q}}{\Delta_{\rm rq}^2},
    \label{Gamma-wf}\end{eqnarray}
where $\kappa_{\rm q}$ is given by Eq.\ (\ref{kappa-q}) and we
assumed  $|\mathcal{G}|^2 \ll  \Delta_{\rm fq}\Delta_{\rm rq}$ to transform Eq.\ (\ref{Gamma-wf-1}) into Eq.\ (\ref{Gamma-wf}).

The Purcell rate given by Eq.\ (\ref{Gamma-wf}) is exactly what we
expected from the classical analysis in Sec.\ II: in the usual
formula (\ref{Gamma-text}) we simply need to substitute $\kappa$
with the readout resonator decay rate $\kappa_{\rm q}$ seen by the qubit.
Since the measurement is governed by a different decay rate
$\kappa_{\rm r}$, the effective Purcell rate suppression factor is
given by Eq.\ (\ref{F-1}), as was expected. This confirms the
results of the classical analysis in Sec.\ II.

\subsection{Method II: Density matrix analysis}

We can also find the Purcell rate in a more traditional way by writing
the master equation \eqref{ma} explicitly in the single-excitation subspace:
\begin{eqnarray}
  && \dot \rho_{\textbf{ee}} = i\textit{g}(\rho_{\textbf{er}}-\rho_{\textbf{re}}), \label{r11} \\
  && \dot \rho_{\textbf{er}} = -i\Delta_{\rm qr} \rho_{\textbf{er}}-i\textit{g}(\rho_{\textbf{rr}}-\rho_{\textbf{ee}})
  +i\mathcal{G}^*\rho_{\textbf{ef}},
  \label{r12} \\
  && \dot \rho_{\textbf{ef}} = -\frac{\kappa_{\rm f}}{2}\rho_{\textbf{ef}}-i\Delta_{\rm qf}\rho_{\textbf{ef}}+i\mathcal{G}\rho_{\textbf{er}} -ig \rho_{\textbf{rf}},
  \label{r13} \\
  && \dot \rho_{\textbf{rr}} =-i\textit{g}(\rho_{\textbf{er}}-\rho_{\textbf{re}})-i\mathcal{G}
  \rho_{\textbf{fr}}+i\mathcal{G}^*\rho_{\textbf{rf}},
  \label{6}\\
    && \dot\rho_{\textbf{rf}}=-\frac{\kappa_{\rm f}}{2}\rho_{\textbf{rf}}+i\Delta_{\rm fr}\rho_{\textbf{rf}}-i\mathcal{G}\rho_{\textbf{ff}}+i \mathcal{G}^* \rho_{\textbf{rr}}
    -ig \rho_{\textbf{ef}}, \qquad \label{7}\\
  &&  \dot \rho_{\textbf{ff}} = -\kappa_{\rm f} \rho_{\textbf{ff}} -i\mathcal{G}^*\rho_{\textbf{rf}}+i \mathcal{G}\rho_{\textbf{fr}}.
   \label{8}
\end{eqnarray}
Note that $\dot \rho_{\textbf{gg}} = \kappa_{\rm f}
\rho_{\textbf{ff}}$ and
$\rho_{\textbf{ee}}+\rho_{\textbf{rr}}+\rho_{\textbf{ff}}+\rho_{\textbf{gg}}=1$ (however, we do not use these two equations for the derivation of the
Purcell rate).

Using the quasisteady solutions of Eqs.\ (\ref{r12})--(\ref{8}),
i.e.\ assuming $\dot \rho_{\textbf{er}}=\dot \rho_{\textbf{ef}}=\dot
\rho_{\textbf{rr}}=\dot \rho_{\textbf{rf}}=\dot
\rho_{\textbf{ff}}=0$, we can obtain a lengthy equation for
$\rho_{\textbf{er}}$, which is proportional to $\rho_{\textbf{ee}}$.
If we use the first-order expansion of this equation in the coupling
$g$ and neglect $g^3$ terms (there is no $g^2$ contribution), then
    \be
    \rho_{\textbf{er}} = \frac{ig}{i\Delta_{\rm rq}+|\mathcal{G}|^2/(i\Delta_{\rm fq}+\kappa_{\rm f}/2)} \, \rho_{\textbf{ee}},
    \ee
which has the form similar to Eq.\ (\ref{lambda-e}). Substituting
this $\rho_{\textbf{er}}$ into Eq.\ (\ref{r11}), we obtain the
evolution equation $\dot\rho_{\textbf{ee}}=-\Gamma
\rho_{\textbf{ee}}$ with $\Gamma$ given exactly by Eq.\
(\ref{Gamma-wf-1}). If we do not use the above-mentioned
approximation for the quasisteady $\rho_{\textbf{er}}$, then the
result for the Purcell rate is slightly different and much
lengthier,
    \begin{eqnarray}
&&    \Gamma = g^2 |\mathcal{G}|^2\kappa_{\rm f}  \big[
    (\Delta_{\rm rq} \Delta_{\rm fq} -|\mathcal G|^2)^2
    + (\Delta_{\rm rq}^2 +g^2)(\kappa_{\rm f}/2)^2
    \nonumber \\
&& \hspace{0.5cm}
    +g^2 (\Delta_{\rm fq}^2 +2\Delta_{\rm fq}\Delta_{\rm rq}
    - |\mathcal G|^2 ) + g^4 \big]^{-1} .
    \label{Gamma-dm}\end{eqnarray}
Thus the derivations based on the wavefunction and density matrix
languages using the quasisteady-state approximation both lead to practically
the same result for the Purcell rate $\Gamma$. The most physically
transparent result is given by Eq.\ (\ref{Gamma-wf}), which
corresponds to the semiclassical analysis in Sec.\ II and simply
replaces $\kappa$ in Eq.\ (\ref{Gamma-text}) with $\kappa_{\rm q}$
seen by the qubit, in contrast to the measurement process, which is
governed by $\kappa_{\rm r}$.

As an example, let us use the parameters similar to that in
Ref.\ \cite{Jeffrey-14}: $\omega_{\rm q}/2\pi = 5.9$ GHz,
$\omega_{\rm r}/2\pi = 6.8$ GHz, $\omega_{\rm f}/2\pi = 6.75$ GHz,
$Q_{\rm f}=30$ (so that $\kappa_{\rm f}^{-1}= 0.71$ ns), $g/2\pi=90$
MHz, and $\kappa_{\rm r}^{-1}=30$ ns (so that $\mathcal{G}/2\pi
=18.9$ MHz). In this case the resonator decay [Eq.\ (\ref{kappa-q})]
seen by the qubit is $\kappa_{\rm q}^{-1}=1.45\, \mu$s, the Purcell
rate [Eq.\ (\ref{Gamma-wf})] is $\Gamma =1/(145\, \mu{\rm s})$, and
the Purcell rate suppression factor [Eq.\ (\ref{F-1})] is
$F=30\,{\rm ns}/1.45\, \mu{\rm s}=(1+0.44^2)/(1+7.6^2)=0.021$.

Thus, for typical parameters the bandpass Purcell filter suppresses
the Purcell decay by a factor of $\sim$50. It is easy to increase
this factor to 100 by using $\omega_{\rm q}/2\pi = 5.5$ GHz in the
above example; however, further decrease of the Purcell rate is not
needed for practical purposes, while increased resonator-qubit
detuning decreases the dispersive shift $2\chi$ (in the above
example $2\chi/2\pi \approx -3$ MHz for the qubit anharmonicity of
180 MHz, while for $\omega_{\rm q}/2\pi = 5.5$ GHz the dispersive
shift becomes twice less).

Note that for the parameters in the above example, Eq.\
(\ref{Gamma-wf}) overestimates the exact solution for $\Gamma$ via
Eq.\ (\ref{lambda-eq}) by 5\%, the same 5\% for Eq.\
(\ref{Gamma-wf-1}), and Eq.\ (\ref{Gamma-dm}) overestimates the
Purcell rate by 2\%. The solution of Eq.\ (\ref{lambda-eq}) as a
quadratic equation neglecting $\lambda^3$ gives $\Gamma$, which
overestimates the exact solution by 22\%, while the second iteration
is practically exact ($-0.01\%$). The inaccuracies grow for smaller
resonator-qubit detuning (crudely as $\Delta_{\rm rq}^{-2}$), but
remain reasonably small in a sufficiently wide range; for example,
Eq.\ (\ref{Gamma-wf}) overestimates the Purcell rate by $50\%$ for
$\omega_{\rm q}/2\pi=6.5$ GHz, i.e.\ detuning of 0.3 GHz.

\subsection{Nonzero readout resonator damping }

In the quantum evolution model (\ref{ma}) we have considered only
the damping of the filter resonator with the rate $\kappa_{\rm f}$.
If there is also an additional energy dissipation in the readout
resonator with the rate $\kappa_{\rm r,d}$ (e.g., due to coupling
with the transmission line delivering the drive $\varepsilon_{\rm
r}$ in Fig.\ \ref{fig2}), then the master equation (\ref{ma}) should
be replaced with
\begin{align}\label{maN}
  \dot \rho=&-i[V,\rho]+\kappa_{\rm f} \left(b\rho b^{\dag}-b^{\dag}b\rho/2-\rho b^{\dag}b/2\right)\notag\\
  &+\kappa_{\rm r,d}(a\rho a^{\dag}-a^{\dag}a\rho/2-\rho a^{\dag}a/2).
\end{align}
In the wavefunction-language derivation this leads to the extra term
$-\kappa_{\rm r,d}c_{\textbf r}/2$ in Eq.\ (\ref{c-dot-r}) for $\dot
c_{\textbf r}$. This does not change the quasisteady value for
$c_{\textbf f}$ but changes the quasisteady value $c_{\textbf
r}=-igc_{\textbf e}/[i\Delta_{\rm rq} + |\mathcal{G}|^2
/(i\Delta_{\rm fq}+\kappa_{\rm f}/2)+\kappa_{\rm r,d}/2]$, so that
the Purcell rate is
    \begin{eqnarray}
&&    \Gamma =2 \, {\rm Re} \left[ \frac{ g^2 } {i\Delta_{\rm rq} +
|\mathcal{G}|^2 /(i\Delta_{\rm fq}+\kappa_{\rm f}/2) +\kappa_{\rm
r,d}/2} \right] \qquad
    \\
    && \hspace{0.5cm} \approx \frac{g^2(\kappa_{\rm q}
    +\kappa_{\rm r,d})}{\Delta_{\rm rq}^2}
    \end{eqnarray}
instead of Eq.\ (\ref{Gamma-wf}). Practically the same result can be obtained
using the derivation via the density matrix evolution (assuming $\kappa_{\rm r, d}\ll \kappa_{\rm f}$). As expected,
the dissipation $\kappa_{\rm r,d}$ simply adds to the rate
$\kappa_{\rm q}$ seen by the qubit. Since $\kappa_{\rm r,d}$ is not
affected by the filter, it adds in the same way to the bandwidth
$\kappa_{\rm r}$ governing the qubit measurement process and thus
deteriorates the Purcell rate suppression (\ref{F-1}), replacing it
with $F=(\kappa_{\rm q}+\kappa_{\rm r,d})/(\kappa_{\rm r}
+\kappa_{\rm r,d})$.

\section{Purcell rate with microwave drive and bandpass filter}

The Purcell rate may decrease when the measurement microwave drive is added \cite{Set14}. A simple physical reason is the ac Stark shift, which in the typical setup increases the absolute value of detuning between the qubit and readout resonator with increasing number of photons in the resonator, thus reducing the Purcell rate. However, this explanation may not necessarily work well quantitatively.

\begin{figure}
\includegraphics[width=8cm]{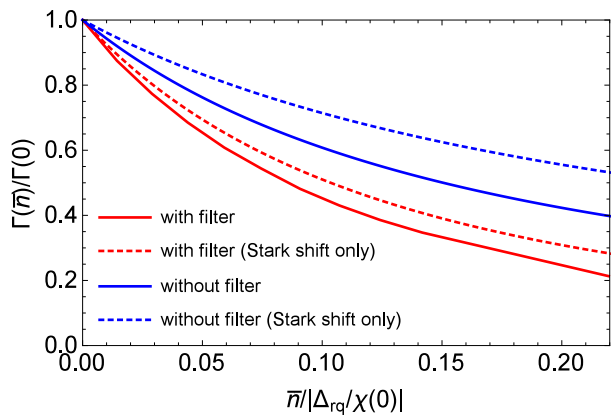}
\caption{The Purcell relaxation rate $\Gamma (\bar{n})$ with a microwave drive,
normalized by the no-drive rate $\Gamma(0)$, with the filter (red solid curve) and without the filter (blue solid curve), as functions of the mean number of photons $\bar n$ in the readout resonator. The numerical simulations used the two-level approximation for the qubit, for which $\bar{n}/|\Delta_{\rm rq}/\chi (0)|=\bar{n}/4n_{\rm crit}$. The dashed lines show the values expected from the model based on the ac Stark shift: $\Gamma ({\bar n})/\Gamma (0)=(1+{\bar n}/n_{\rm crit})^{-2}$ with the filter and $(1+{\bar n}/n_{\rm crit})^{-1}$ without the filter. The parameters used in the simulations are given in the text. }\label{fig4}
\end{figure}

The Purcell rate suppression due to the microwave drive was analyzed
in Ref.\ \cite{Set14} for the case without the Purcell filter and
using the two-level approximation for the qubit. It was shown that
in this case the suppression factor $\Gamma (\bar{n})/\Gamma (0)$ is
approximately $[(1+\bar{n}/n_{\rm crit})^{-1/2}+(1+\bar{n}/n_{\rm
crit})^{-1}]^2$ instead of the factor $(1+\bar{n}/n_{\rm
crit})^{-1}$ expected from the ac Stark shift, where $\bar{n}$ is
the mean number of photons in the resonator and $n_{\rm crit}\equiv
(\Delta_{\rm rq}/2g)^2$. This difference results in the ratio $3/2$
between the corresponding slopes of $\Gamma (\bar{n})$ at small
$\bar{n}$, with the ac Stark shift model underestimating the Purcell
rate suppression (see the blue lines in Fig.\ \ref{fig4}). However, when the third level of the qubit is taken into account, then the ac Stark shift model describes correctly the slope of $\Gamma (\bar{n})$ at small $\bar{n}$  when the qubit anharmonicity is relatively small, $|\delta_{\rm q}/\Delta_{\rm rq}| \ll 1$ (see Appendix). In this case the ac Stark shift model predicts $\Gamma
(\bar{n})/\Gamma (0)=1+4 \bar{n}\chi (0) /\Delta_{\rm rq}$ at
$\bar{n}\ll n_{\rm crit}$, where $\chi (0)$ is the value of $\chi$ at $\bar{n}=0$; note that $|\chi (0)| \ll |g^2/\Delta_{\rm rq}|$ and $\chi (0) <0$ when
$\Delta_{\rm rq}> 0$.

With the filter resonator, we also expect that the ac Stark shift
model for the Purcell rate suppression should work reasonably well,
so that from Eq.\ (\ref{Gamma-wf}) we expect
    \be
  \Gamma (\bar{n}) \approx \frac{g^2 |\mathcal{G}|^2\kappa_{\rm f}}
 {[\omega_{\rm r}-\omega_{\rm q,eff} (\bar{n})]^2
 \{[\omega_{\rm f}-\omega_{\rm q,eff} (\bar{n})]^2 +(\kappa_{\rm f}/2)^2\} },
    \ee
where $\omega_{\rm q,eff} (\bar{n})$ is the effective qubit frequency, $\omega_{\rm q,eff} (\bar{n})=\omega_{\rm q}^{\rm b}+2\chi (0) \bar{n}$ if
we neglect dependence of $\chi$ on $\bar{n}$ and the ``Lamb shift''.
Therefore, in a typical situation when $|\omega_{\rm f} - \omega_{\rm r}| \ll |\omega_{\rm r} - \omega_{\rm q}|$ and $\kappa_{\rm f} \ll |\omega_{\rm r} - \omega_{\rm q}|$, we expect the suppression ratio
    \be
    \frac{\Gamma (\bar{n})}{\Gamma (0)} \approx
    \left[\frac{\omega_{\rm r}-\omega_{\rm q}^{\rm b}}
    {\omega_{\rm r}-\omega_{\rm q,eff} (\bar{n})} \right]^4 .
    \ee

To check the accuracy of this formula numerically, we need to add into the Hamiltonian (\ref{Ham-start}) the terms describing the drive and higher levels in the qubit (see Appendix). However, the resulting Hilbert space was too large for our numerical simulations, so we numerically calculated $\Gamma (\bar{n})$ using only the two-level approximation for the qubit. Using the rotating frame based on the drive frequency $\omega_{\rm d}$ [i.e., $H_0=\omega_{\rm d}(\sigma_+\sigma_- +a^\dagger a+ b^\dagger b)$], we then obtain the interaction Hamiltonian
\begin{align}\label{h}
  V_{\rm d}&=\Delta_{\rm rd}a^{\dag}a+\Delta_{\rm fd}b^{\dag}b+ \Delta_{\rm qd} \sigma_+\sigma_- +\textit{g}(a^{\dag}\sigma_{-}+a\sigma_{+})\notag\\
  &+\mathcal{G}a^{\dag}b+ \mathcal{G}^*ab^{\dag} +\varepsilon_{\rm r} a^{\dag}
  + \varepsilon_{\rm r}^* a,
  \end{align}
where $\Delta_{\rm rd}=\omega_{\rm r}^{\rm b}-\omega_{\rm d}$,  $\Delta_{\rm fd}=\omega_{\rm f}-\omega_{\rm d}$, $\Delta_{\rm qd}=\omega_{\rm q}^{\rm b}-\omega_{\rm d}$, and for simplicity we assumed that the drive $\varepsilon_{\rm r}$
is applied to the readout resonator. (For the rotating wave approximation we also need to assume that $|\Delta_{\rm rd}|$, $|\Delta_{\rm fd}|$, $|\Delta_{\rm qd}|$, $|g|$, $|\mathcal{G}|$, and $|\dot{\varepsilon_{\rm r}}/\varepsilon_{\rm r}|$ are all small compared with $\omega_{\rm d}$.) Note that in the two-level
approximation $[\omega_{\rm r}-\omega_{\rm q,eff} (\bar{n}) ]^2
=\Delta_{\rm rq}^2+4g^2\bar{n}=\Delta_{\rm rq}^2 (1+\bar{n}/n_{\rm
crit})$.

We have numerically solved the full master equation with the
Hamiltonian (\ref{h}), including the decay $\kappa_{\rm f}$ of the
filter resonator. As the initial state we use the excited state for
the qubit and vacuum for the two resonators, $|\psi\rangle_{\rm
in}=|e00\rangle$. The Purcell rate is extracted from the numerical
solution of $\rho_{ee}(t)$ by fitting $-\ln[\rho_{ee}(t)]$ with a
linear function in the long time limit (still requiring $1-\rho_{ee}(t)\ll 1$). In the simulations we pump
the readout resonator with the frequency $\omega_{\rm d}=\omega_{\rm
r}^{|e\rangle}$ and control $\bar{n}$ by
choosing the corresponding value of $\varepsilon_{\rm r}$. The value of $\bar{n}$ is calculated numerically; it is close to what is expected from the solution of the classical field equations when the $n$-dependence of $\chi$ and $\omega_{\rm r}^{|e\rangle}$ is taken into account: in the two-level approximation $\chi({\bar n})=-g^2/[\Delta_{\rm rq}\sqrt{1+4g^2{\bar n}/\Delta_{\rm rq}^2}]$ and $\omega_{\rm r}^{|e\rangle}=\omega_{\rm r}^{\rm b}+\chi ({\bar n})$. Since $\omega_{\rm r}^{|e\rangle}$ changes with $\bar n$, we change $\omega_{\rm d}$ accordingly.

The red solid line in Fig.\  \ref{fig4} shows the numerical results for the Purcell rate suppression factor $\Gamma (\bar{n})/\Gamma (0)$ as a function of $\bar n=\bar{n}^{|e\rangle}$, normalized by $|\Delta_{\rm rq}/\chi(0)|$. Note that in the two-level approximation (which we used in the simulations) ${\bar n}/|\Delta_{\rm rq}/\chi(0)|={\bar n}/4n_{\rm crit}$. In the simulations we have used $g/2\pi=100\,$MHz, $\kappa_{\rm r}^{-1}=36\,$ns, $\kappa_{\rm f}^{-1}=0.71\,$ns, $\omega_{\rm r}^{\rm b}/2\pi=\omega_{\rm f}/2\pi=6.8\,$GHz,
and $\omega_{\rm q}^{\rm b}/2\pi=6\,$GHz. The blue solid line shows the numerical suppression factor for the standard setup \cite{Set14} (without the filter resonator), also in the two-level approximation for the qubit. We see a larger suppression for the case with the filter, as expected from
the ac Stark shift interpretation and the fact that $\Gamma \propto \Delta_{\rm rq}^{-4}$ with the filter, while $\Gamma \propto \Delta_{\rm rq}^{-2}$ in the standard setup.
However, comparison with the prediction of the ac Stark shift model (dashed lines) does not show a quantitative agreement. There is about 10\% discrepancy for the slope of $\Gamma (\bar{n})$  between the red solid and red dashed lines in the case with the filter. This is a better agreement than for the case without the filter (blue lines).

Note that the numerical calculations have been made using only the two-level model for the qubit. It is possible that the agreement between the ac Stark shift model and the numerical results is much better if three or more levels in the qubit are taken into account. This is an interesting question for further research.

Also note that in experiments, increase of the drive power often leads to decrease of the qubit lifetime, instead of the increase, predicted by our analysis (both with and without the filter). The reason for this effect is not quite clear and may be related to various technical issues. Therefore, either suppression or enhancement of the qubit relaxation with the drive power may be observed in actual experiments.

\section{Conclusion}

In this paper we have discussed the theory of the bandpass Purcell filter used in Refs.\ \cite{Jeffrey-14,Kelly-15} for measurement of superconducting qubits. An additional wide-bandwidth filter resonator (Fig.\ \ref{fig2}) coupled to the readout resonator easily passes the microwave field used for the qubit measurement, but strongly impedes the propagation of a photon at the qubit frequency, which is far outside of the filter bandwidth. A simple way to quantitatively describe the operation of the filter is by noticing that the effective decay rate $\kappa_{\rm eff}$ of the readout resonator [Eq.\ (\ref{kappa-eff})] depends on frequency. Therefore, the measurement is governed by a relatively large value $\kappa_{\rm r}$ [Eq.\ (\ref{kappa-r})], which permits a fast measurement, while the Purcell relaxation is determined by a much smaller value $\kappa_{\rm q}$ seen by the qubit [Eq.\ (\ref{kappa-q})]. The ratio of these effective decay rates of the readout resonator gives the suppression factor for the qubit relaxation [Eq.\ (\ref{F-1})]. The result for the suppression factor is similar to the result obtained in Ref.\ \cite{Jeffrey-14} using circuit theory (with a few minor differences).

We have first analyzed the operation of the Purcell filter quasiclassically, and then confirmed the results using the quantum approach. In the quantum analysis we have used two approaches: based on decaying wavefunction and density matrix evolutions. While the Purcell effect is traditionally described using density matrices, it is actually simpler to use the approach based on wavefunctions. The results of our semiclassical and two quantum approaches are very close to each other; however, they are not identical because the approaches use slightly different approximations. A simple and most physically transparent result for the Purcell rate is given by Eq.\ (\ref{Gamma-wf}).

The Purcell rate of the qubit decay is further suppressed when a microwave drive is applied for measurement. The effect is similar to what was discussed in Ref.\ \cite{Set14} for the standard setup without the filter, and can be crudely understood as being due to the ac Stark shift of the qubit frequency, which increases the resonator-qubit detuning $\Delta_{\rm rq}$. Since the Purcell rate with the filter scales with $\Delta_{\rm rq}$ crudely as $\Delta_{\rm rq}^{-4}$   [Eq.\ (\ref{Gamma-wf})] instead of $\Delta_{\rm rq}^{-2}$ in the standard setup, the Purcell rate suppression due to microwaves is stronger in the case with the filter. Numerical results for the suppression due to microwaves using the two-level approximation for the qubit show that the explanation based on ac Stark shift works well, but underestimates the effect by about 10\%. This discrepancy may be significantly less if more levels in the qubit are taken into account; however, this still remains an open question.

The bandpass Purcell filter decreases the qubit decay due to the Purcell effect by a factor of $\sim$50 for typical parameters, for the same measurement conditions as in the standard setup without the filter. This allows much faster and more accurate measurement of superconducting qubits, compared to the case without the filter. The qubit measurement within 140 ns with 99\% fidelity using this filter has been demonstrated in Ref.\ \cite{Jeffrey-14}. With a slight change of parameters it seems possible to perform qubit readout within $\sim$50 ns with fidelity approaching 99.9\%. Such fast and accurate qubit readout would be very useful for quantum information processing with superconducting qubits.

\begin{acknowledgments}
The authors thank Daniel Sank,  Mostafa Khezri, and Justin Dressel for useful discussions.
  The research was funded by the
Office of the Director of National Intelligence (ODNI), Intelligence
Advanced Research Projects Activity (IARPA), through the Army
Research Office Grant No. W911NF-10-1-0334. All statements of fact,
opinion or conclusions contained herein are those of the authors and
should not be construed as representing the official views or
policies of IARPA, the ODNI, or the U.S. Government. We also
acknowledge support from the ARO MURI Grant No. W911NF-11-1-0268.
\end{acknowledgments}

\appendix
\section{Qubit measurement and Purcell effect in the standard setup }

In this Appendix we review the trade-off between the Purcell rate
and measurement time for a transmon or Xmon qubit in the standard
cQED setup (without a filter).  In the standard setup
\cite{Blais-04,Wallraff-04,Koch-07} (Fig.\ \ref{fig1}) the qubit is
dispersively coupled with the resonator, so that the qubit state
slightly changes the resonator frequency. This change causes a phase
shift (and in general an amplitude change) of a microwave field
transmitted through or reflected from the resonator. The transmitted
or reflected microwave is then amplified and sent to a mixer, so
that the phase shift (and amplitude change) can be discriminated,
thus distinguishing the states $|g\rangle$ and $|e\rangle$ of the
qubit.

\subsection{Basic theory}

For the basic analysis of measurement \cite{Koch-07}, it is
sufficient to consider {\it three} energy levels of the qubit: the ground
state $|g\rangle$, the first excited state $|e\rangle$, and the
second excited state $|f\rangle$, so that the Hamiltonian is
($\hbar=1$)
    \begin{align}
    H & =  \omega_{\rm q}^{\rm b} |e\rangle \langle e| +
    (2\omega_{\rm q}^{\rm b}-\delta_{\rm q}) |f\rangle \langle f| +
    \omega_{\rm r}^{\rm b} a^{\dag}a
    \nonumber \\
& + g a^{\dag}|g\rangle\langle e| +g^* a |g\rangle\langle e|
 + \tilde{g} a^{\dag}|e\rangle\langle f| +
 \tilde{g}^* a |f\rangle\langle e|
    \notag\\
&+\varepsilon_{\rm r} a^\dagger e^{-i\omega_{\rm d}t}+\varepsilon_{\rm r}^*
a e^{i\omega_{\rm d}t}+H_{\kappa}+H_{\gamma},
    \label{H-standadr-init}\end{align}
where $\omega_{\rm q}^{\rm b}$ is the bare qubit frequency,
$\delta_{\rm q}$ is its ahnarmononicity ($\delta_{\rm q}>0$,
$\delta_{\rm q}/ \omega_{\rm q}^{\rm b} \sim 0.2\, {\rm GHz}/6\,
{\rm GHz}\ll 1$), $\omega_{\rm r}^{\rm b}$ is the bare resonator
frequency, $a$ is the annihilation operator for the resonator, $g$
is the coupling between the qubit and the resonator,
$\tilde{g}\approx \sqrt{2}\,g$ is the similar coupling involving
levels $|e\rangle$ and $|f\rangle$, and $\varepsilon_{\rm r}$ is the
normalized amplitude of the microwave drive with frequency
$\omega_{\rm d}$. For brevity $H_{\kappa}$ describes the coupling of
the resonator with the transmission line, which causes resonator
energy damping with the rate $\kappa$, while $H_\gamma$ describes
the intrinsic qubit relaxation (excluding the Purcell effect) with
the rate $T_{1,\rm int}^{-1}$. Note that in the Hamiltonian
(\ref{H-standadr-init}) we neglected the coupling terms creating or
annihilating the double excitations in the qubit and the resonator.
For simplicity we assume a real coupling: $g^*=g$ and
$\tilde{g}^*=\tilde{g}$.

For a simple analysis of the measurement process, let us first
neglect $H_\kappa$, $H_\gamma$, and the drive $\varepsilon_r$, and
consider the three Jaynes-Cummings ladders of states
$|g,n\rangle$, $|e,n\rangle$, and $|f,n\rangle$, where $n$ denotes
the number of photons in the resonator. The coupling $g$ provides
the level repulsion between $|g,n+1\rangle$ and $|e,n\rangle$
(effective coupling is $\sqrt{n+1}\, g$), while $\tilde{g}$ provides
the level repulsion between $|e,n+1\rangle$ and $|f,n\rangle$ (with
coupling $\sqrt{n+1}\, \tilde{g}$). Assuming sufficiently large level
separation, $|\omega_{\rm q}^{\rm b}-\omega_{\rm r}^{\rm b} |\gg
\sqrt{n} \, |g|$ and $|\omega_{\rm q}^{\rm b}-\delta_{\rm q}
-\omega_{\rm r}^{\rm b}| \gg \sqrt{n} \, |\tilde{g}|$, we can treat
the level repulsion to lowest order; then the energies of the
eigenstates $\overline{|g,n\rangle}$ and $\overline{|e,n\rangle}$
are
    \begin{align}
&    E_{\overline{|g,n\rangle}}= n\omega_{\rm r}^{\rm b} -  \,
\frac{n g^2}{\omega_{\rm q}^{\rm b}-\omega_{\rm r}^{\rm b}},
    \label{eigenenergy-g}\\
 & E_{\overline{|e,n\rangle}}=
\omega_{\rm q}^{\rm b }+ n\omega_{\rm r}^{\rm b} + \, \frac{(n+1)\,
g^2}{\omega_{\rm q}^{\rm b}-\omega_{\rm r}^{\rm b}} -
\frac{n\tilde{g}^2}{\omega_{\rm q}^{\rm b} -\delta_{\rm
q}-\omega_{\rm r}^{\rm b}}.
    \label{eigenenergy-e}\end{align}
Therefore the effective resonator frequency $\omega_{\rm
r}^{|g\rangle}$ when the qubit in the ground state is
    \be
  \omega_{\rm r}^{|g\rangle} = E_{\overline{|g,n+1\rangle}} - E_{\overline{|g,n\rangle}}
  = \omega_{\rm r}^{\rm b} - \frac{g^2}{\Delta},
    \label{omega-r-g}\ee
where
    \be
    \Delta =\Delta_{\rm qr}=\omega_{\rm q}^{\rm b}-\omega_{\rm
  r}^{\rm b},
    \label{Delta-def}\ee
while for the qubit state $|e\rangle$ the effective resonator
frequency is
    \be
  \omega_{\rm r}^{|e\rangle} = E_{\overline{|e,n+1\rangle}} \, -
  E_{\overline{|e,n\rangle}}
  = \omega_{\rm r}^{\rm b} + \frac{g^2}{\Delta} -
  \frac{\tilde{g}^2}{\Delta - \delta_{\rm q}}.
    \label{omega-r-e}\ee
Denoting the frequency difference by $2\chi$, we obtain
    \begin{eqnarray}
&& \hspace{2cm}  \omega_{\rm r}^{|e\rangle}-  \omega_{\rm r}^{|g\rangle} = 2\chi,
 \,\,\,
    \label{chi-def}\\
&&  \chi = \frac{g^2}{\Delta} -
   \frac{\tilde{g}^2/2}{\Delta-\delta_{\rm q}} =-\frac{g^2\delta_{\rm q}}{\Delta (\Delta-\delta_{\rm q})} + \frac{g^2-\tilde{g}^2/2}{\Delta-\delta_{\rm q}}. \qquad
    \label{chi-res}\end{eqnarray}
The corresponding effective qubit frequency is
    \be
    \omega_{\rm q}^{\rm eff}  = E_{\overline{|e,n\rangle}} -
  E_{\overline{|g,n\rangle}}
  = \omega_{\rm q}^{\rm b }
  + \, \frac{g^2}{\Delta} +2n\chi,
    \label{omega-q-eff}\ee
which includes the ``Lamb shift'' $g^2/\Delta$ and the ``ac Stark
shift'' $2n\chi$.

Therefore, in this case the first two lines of the Hamiltonian
(\ref{H-standadr-init}) can be approximated as
    \be
 H = \frac{\omega_{\rm q}}{2} \, \sigma_z + \omega_{\rm r} a^{\dag}a +
 \chi a^{\dag}a \, \sigma_z,
   \label{H-disp}\ee
where
    \be
\omega_{\rm q}=\omega_{\rm q}^{\rm b}+\frac{g^2}{\Delta}, \,\,\,\,
\omega_{\rm r}= \omega_{\rm r}^{\rm b}- \frac{1}{2}\,
\frac{\tilde{g}^2}{\Delta - \delta_{\rm q}},
    \label{omega-q-r}\ee
$\sigma_z=|e\rangle\langle e|-|g\rangle\langle g|$, we shifted the
energy by $-\omega_{\rm q}/2$, and we no longer need the qubit state
$|f\rangle$.

Note that the dispersive coupling $\chi$ given by Eq.\
(\ref{chi-res}) is much smaller \cite{Koch-07} than the value
$g^2/\Delta$ expected in the two-level case. This is because the
transmon and Xmon qubits are only slightly different from a linear
oscillator, for which $\tilde{g}=\sqrt{2}\, g$ and $\delta_{\rm
q}=0$, thus leading to $\chi=0$. The effect of nonzero anharmonicity $\delta_{\rm
q}$ in Eq.\ (\ref{chi-res}) is more important than the effect of nonzero
$\tilde{g}-\sqrt{2}\, g$ because
    \be
\frac{\tilde{g}^2}{2}-g^2 \approx - g^2 \frac{\delta_{\rm
q}}{\omega_{\rm q}^{\rm b}} \,\,\,
    \label{g-tilde}\ee
 and $|\Delta|\ll
\omega_{\rm q}^{\rm b}$. Therefore $\chi$ can be approximated as
    \be
    \chi \approx -\frac{g^2\delta_{\rm q}}{\Delta (\Delta-\delta_{\rm q})} \approx
    - \frac{g^2 \delta_{\rm q}}{\Delta^2},
    \label{chi-approx}\ee
where the last formula also assumes $|\Delta| \gg \delta_{\rm q}$.

The dispersive approximation (\ref{H-disp}) is based on the
approximate formulas (\ref{eigenenergy-g})--(\ref{eigenenergy-e})
for the eigenenergies, and therefore requires a limited number of
photons $n$ in the resonator,
    \be
    n \ll \min (n_{\rm crit}, \tilde{n}_{\rm crit}), \,\,\,
    n_{\rm crit} =\frac{\Delta^2}{4g^2}, \,\,\,
    \tilde{n}_{\rm crit} =\frac{(\Delta-\delta_{\rm q})^2}{4\tilde{g}^2},
    \label{n<<n-crit}\ee
where the factor 4 in the definitions of the critical photon numbers
$n_{\rm crit}$ and $\tilde{n}_{\rm crit}$ is a usual convention. For
$n$ beyond this range it is still possible (at least to some extent)
to use the dispersive approximation (\ref{H-disp}) if the spread of
$n$ is relatively small; however $\omega_{\rm q}$, $\omega_{\rm r}$,
and $\chi$ should be redefined. For that we need to use similar
steps as in Eqs.\ (\ref{omega-r-g})--(\ref{omega-q-eff}), but
starting with more accurate formulas than (\ref{eigenenergy-g}) and
(\ref{eigenenergy-e}). In particular, in the two-level approximation
for the qubit ($\tilde{g}=0$ or $\delta_{\rm q}=\infty$) we would
obtain $\chi \approx g^2/\sqrt{\Delta^2+4\bar{n}g^2}$ with $\bar{n}$
being the average (typical) number of photons; however,
the generalization of the realistic case (\ref{chi-approx}) is not so
simple (see below).

The dispersive approximation (\ref{H-disp}) cannot reproduce the
Purcell effect \cite{Blais-04} after including the last line of the
Hamiltonian (\ref{H-standadr-init}), so it should be added
separately. Without the microwave drive ($\varepsilon_{\rm r}=0$)
only levels $|e,0\rangle$, $|g,1\rangle$ and $|g,0\rangle$ are
involved into the evolution described by Eq.\
(\ref{H-standadr-init}), and for sufficiently small resonator
bandwidth, $\kappa \ll \sqrt{\Delta^2+4g^2}$, the qubit relaxation
rate due to Purcell effect is
    \be
    \Gamma \approx \frac{\kappa}{2} \left( 1-
    \frac{|\Delta|}{\sqrt{\Delta^2+4g^2}} \right) \approx
    \frac{g^2 \kappa}{\Delta^2}.
    \label{Gamma-no-drive}\ee
Note that this rate does not depend on the qubit anharmonicity
$\delta_{\rm q}$, in contrast to the dispersive coupling $\chi$,
which vanishes at $\delta_{\rm q} \rightarrow 0$. This is because
the {\it Purcell effect is essentially a linear effect} (energy decay via
decay of a coupled system), in contrast to $\chi$, which is based on
qubit nonlinearity. This linearity is the reason why the Purcell
rate $\Gamma$ does not change (in the first approximation) when the
microwave drive $\varepsilon_{\rm r}$ creates a significant photon
population in the resonator, $1\ll n \ll \min (n_{\rm crit},
\tilde{n}_{\rm crit})$. (Someone might naively expect that $\Gamma$
scales with $n$ because of effective coupling $\sqrt{n}\, g$.)
However, in the next approximation $\Gamma$ depends on $n$
\cite{Set14} and can be calculated as
    \be
    \Gamma (n) = \kappa \, |\overline{\langle g, n|} a
    \overline{|e,n\rangle }|^2,
    \label{Gamma-n-def}\ee
with subsequent replacement of $n$ with the average photon number
$\bar{n}$ if the spread of $n$ is relatively small.

    Using the third-order (in $g$ and/or $\tilde{g}$) eigenstates,
    \begin{align}
  &    \overline{|g,n\rangle } = \left( 1-\frac{n
g^2}{2\Delta^2}\right) |g,n\rangle +\frac{\sqrt{n(n-1)}\, g
\tilde{g}}{\Delta(2\Delta-\delta_{\rm q})} \, |f,n-2\rangle
    \nonumber \\
& \hspace{0.5cm}   -\frac{\sqrt{n}\, g}{\Delta} \left( 1-\frac{3n g^2}{2\Delta^2} + \frac{(n-1)\tilde{g}^2}{\Delta(2\Delta-\delta_{\rm q})}\right)  |e,n-1\rangle ,
    \label{g,n-eigen}\\
& \overline{|e,n\rangle } \approx \left( 1-\frac{(n+1) g^2}{2\Delta^2}
    -\frac{n \tilde{g}^2}{2(\Delta-\delta_{\rm q})^2}\right) |e,n\rangle
     \nonumber \\
    & \hspace{0.5cm} + \frac{\sqrt{n+1}\, g}{\Delta} \left( 1- \frac{3(n+1) g^2}{2\Delta^2} +\frac{n\tilde{g}^2 (\Delta-2\delta_{\rm q})}{2\Delta (\Delta-\delta_{\rm q})^2} \right)
    \nonumber \\
    & \hspace{0.5cm} \times  |g,n+1\rangle
    -  \frac{\sqrt{n}\, \tilde{g}}{\Delta-\delta_{\rm q}} \,
    |f,n-1\rangle ,
    \label{e,n-eigen}\end{align}
where the last term in Eq.\ (\ref{e,n-eigen}) for brevity is only of the first order, we find the Purcell rate
    \be
    \Gamma (n) = \kappa \frac{g^2}{\Delta^2} \left(1- \frac{3g^2}{\Delta^2}
    -6 n \frac{g^2}{\Delta^2} +n \frac{\tilde{g}^2 (3\Delta -4\delta_{\rm q})}
    {\Delta (\Delta-\delta_{\rm q})^2} \right)  ,
    \label{Gamma-n-1}\ee
which is an approximation up to 5th order in $g$. This result
coincides with the result of Ref.\ \cite{Set14} when $\tilde{g}=0$.
Approximating $\tilde{g}=\sqrt{2}\, g$ [see Eq.\
(\ref{g-tilde})], we obtain
    \be
  \Gamma (n) \approx \kappa \frac{g^2}{\Delta^2} \left(
  1- \frac{3g^2}{\Delta^2}
    + n \, \frac{2g^2 \delta_{\rm q} (2\Delta -3\delta_{\rm q})}
    {\Delta^2 (\Delta-\delta_{\rm q})^2} \right) .
    \label{Gamma-n-2} \ee
It is interesting to note that while in the two-level approximation
($\tilde{g}=0$) the Purcell rate is always suppressed with increasing
$n$ \cite{Set14}, Eq.\ (\ref{Gamma-n-2}) shows the suppression only
when $\Delta < (3/2)\delta_{\rm q}$ (which is the usual experimental
case, since $\Delta$ is usually negative). Numerical results using
Eq.\ (\ref{Gamma-n-def}) show that even when $\Gamma(n)$ initially
increases with $n$, it is still suppressed at larger $n$. Note that the result (\ref{Gamma-n-2}) requires
assumption (\ref{n<<n-crit}) of a sufficiently small nonlinearity.
Comparing Eqs.\ (\ref{Gamma-n-2}) and (\ref{chi-approx}), we see
that in the case of large detuning, $|\Delta|\gg \delta_{\rm q}$,
the dependence of the Purcell rate on $n$ is consistent with the
explanation based on the ac Stark shift, $\Gamma (n)\approx \kappa
g^2/ (\Delta +2n \chi)^2$. (This is in contrast to the two-level
approximation, in which this explanation leads to a discrepancy
in the slope \cite{Set14} by a factor of $3/2$.) We have checked that taking into account the fourth level in the qubit does not change Eqs.\ (\ref{Gamma-n-1}) and (\ref{Gamma-n-2}); there are no additional contributions of the order $g^4$.

Besides the qubit energy relaxation $\Gamma$, the resonator damping
$\kappa$ in the presence of drive leads to the qubit excitation \cite{Set14}
$|g\rangle \rightarrow |e\rangle$ with the rate $\Gamma_{|g\rangle \rightarrow |e\rangle}=\kappa \,
|\overline{\langle e, n-2|} a \overline{|g,n\rangle }|^2$ and the
excitation $|e\rangle \rightarrow |f\rangle$ with the rate $\Gamma_{|e\rangle \rightarrow |f\rangle}=\kappa
\, |\overline{\langle f, n-2|} a \overline{|e,n\rangle }|^2$. These
rates (up to sixth order in coupling) are
\begin{eqnarray}
&& \Gamma_{|g\rangle\rightarrow|e\rangle}= \frac{\kappa g^2n(n-1)}{\Delta^2}\left[\frac{g^2}{\Delta^2}-\frac{\tilde g^2}{\Delta(2\Delta-\delta_{\rm q})}\right]^2 \qquad
    \\
&& \hspace{1.3cm} \approx  \frac{\kappa g^2}{\Delta^2}\left( \frac{n}{n_{\rm crit}}\right )^2\frac{\delta_{\rm q}^2}{16 \, (2\Delta-\delta_{\rm q})^2} ,
 \label{Gamma-g-e-2}\end{eqnarray}
\begin{eqnarray}
&& \hspace{-0.0cm} \Gamma_{|e\rangle\rightarrow|f\rangle}= \frac{\kappa \tilde g^2n(n-1)}{(\Delta-\delta_{\rm q})^4}\left[ \frac{\tilde g^2}{\Delta-\delta_{\rm q}}-\frac{g^2}{2\Delta-\delta_{\rm q}} \right.  \,\,\,
    \nonumber \\
&& \hspace{1.4cm}  \left. -\frac{\tilde{\tilde g}^2}{2\Delta-3\delta_{\rm q}}\right]^2
    \\
&&   \hspace{0.8cm} \approx  \frac{\kappa g^2}{\Delta^2}\frac{ ( n/n_{\rm crit})^2 \, \delta_{\rm q}^2 \Delta^8}{ 2(\Delta-\delta_{\rm q})^6(2\Delta-\delta_{\rm q})^2(2\Delta-3\delta_{\rm q})^2},  \qquad \,\,\,\,
   \label{Gamma-e-f-2}\end{eqnarray}
where $\tilde{\tilde g}\approx \sqrt{3}\, g$ is the coupling due to the fourth qubit level with energy $3\omega_{\rm q}^{\rm b}-3\delta_{\rm q}$, and in Eqs.\ (\ref{Gamma-g-e-2}) and (\ref{Gamma-e-f-2}) we also used $\tilde g=\sqrt{2}\, g$  and replaced $n(n-1)$ with $n^2$.
In the assumed range, $n\ll n_{\rm crit}$, the excitation rates are much smaller than the Purcell decay rate $\Gamma$.

Now let us discuss the $n$-dependence of the dispersive coupling $\chi(n)=[\omega_{\rm r}^{| e\rangle}(n)-\omega_{\rm r}^{|g\rangle}(n)]/2$, where $\omega_{\rm r}^{| g\rangle}(n)=E_{\overline{|g,n+1\rangle}} \, - E_{\overline{|g,n\rangle}}$ and $\omega_{\rm r}^{| e\rangle}(n)=E_{\overline{|e,n+1\rangle}} \, - E_{\overline{|e,n\rangle}}$ [see Eqs.\ (\ref{omega-r-g}), (\ref{omega-r-e}), and (\ref{chi-def})]. Using the three-level approximation for the qubit with accuracy up to the fourth order in $g$ and/or $\tilde g$ (assuming $n\ll n_{\rm crit}$), we obtain
    \begin{eqnarray}
&&    \omega_{\rm r}^{|g\rangle}(n)= -\frac{g^2}{\Delta }+\frac{g^4 (2 n+1)}{\Delta ^3}-\frac{2 g^2 \tilde g^2 n}{\Delta ^2 (2 \Delta -\delta_{\rm q} )},
    \label{omega-r-g-3lev}\\
&& \omega_{\rm r}^{|e\rangle}(n) = \frac{g^2}{\Delta }
-\frac{\tilde g^2}{\Delta -\delta_{\rm q} }
-\frac{g^4 (2 n+3)}{\Delta ^3}
+\frac{\tilde g^4 (2 n+1)}{(\Delta -\delta_{\rm q} )^3} \qquad
    \nonumber \\
&& \hspace{1.3cm}    -\frac{2 \delta_{\rm q}  g^2 \tilde g^2 (n+1)}{\Delta ^2 (\Delta -\delta_{\rm q} )^2},
     \label{omega-r-e-3lev}\end{eqnarray}
so that assuming $\tilde{g}=\sqrt{2}\, g$, we obtain
    \begin{eqnarray}
&&    \chi(n)\approx -\frac{g^2\delta_{\rm q}}{\Delta (\Delta-\delta_{\rm q})}
+\frac{4g^4\delta_{\rm q}}{\Delta^4} \, \frac{1-\tilde\delta_{\rm q}+\tilde\delta_{\rm q}^2/2}{(1-\tilde\delta_{\rm q})^3}
    \nonumber \\
 && \hspace{1.2cm} +
    \frac{3ng^4}{\Delta^3} \, \frac{1-\tilde\delta_{\rm q}^2+\tilde\delta_{\rm q}^3-\tilde\delta_{\rm q}^4/3}{(1-\tilde\delta_{\rm q}/2)(1-\tilde\delta_{\rm q})^3},
    \label{chi(n)-3}\end{eqnarray}
where $\tilde\delta_{\rm q}=\delta_{\rm q}/\Delta$. This result predicts a quite strong $n$-dependence of $\chi$, which is, however, not correct. The reason is that it is not sufficient to consider three qubit levels for $\chi(n)$. When the
fourth level of the qubit is taken into account (with energy $3\omega_{\rm q}^{\rm b}-3\delta_{\rm q}$ and coupling $\tilde{\tilde g}$), this does not change Eq.\ (\ref{omega-r-g-3lev}) for $\omega_{\rm r}^{|g\rangle}(n)$, but introduces an additional term
    \be
    -\frac{2\tilde g^2\tilde{\tilde g}^2 n}{(\Delta-\delta_{\rm q})^2(2\Delta-3\delta_{\rm q})}
    \label{fourth-level-correction}\ee
into Eq.\ (\ref{omega-r-e-3lev}) for $\omega_{\rm r}^{|e\rangle}(n)$. Assuming $\tilde{\tilde g}= \sqrt{3}\, g$, this changes Eq.\ (\ref{chi(n)-3}) into
        \begin{eqnarray}
&&    \chi(n)\approx -\frac{g^2\delta_{\rm q}}{\Delta (\Delta-\delta_{\rm q})}
+\frac{4g^4\delta_{\rm q}}{\Delta^4} \, \frac{1-\tilde\delta_{\rm q}+\tilde\delta_{\rm q}^2/2}{(1-\tilde\delta_{\rm q})^3}
    \nonumber \\
 &&  \hspace{0.5cm} -
    \frac{9ng^4\delta_{\rm q}^2}{2\Delta^5} \, \frac{1-(5/3)\tilde\delta_{\rm q}+(11/9)\tilde\delta_{\rm q}^2-\tilde\delta_{\rm q}^3/3}{(1-\tilde\delta_{\rm q}/2)(1-\tilde\delta_{\rm q})^3(1-3\tilde\delta_{\rm q}/2)}. \qquad \,\,\,
    \label{chi(n)-4}\end{eqnarray}
which shows a quite weak dependence on $n$,
    \be
\frac{d\chi (n)}{dn} \approx \frac{9}{8}\, \frac{\delta_{\rm q}}{\Delta } \, \frac{\chi (0)}{n_{\rm crit}},
    \label{chi-slope}\ee
assuming $\delta_{\rm q}\ll |\Delta|$. In the usual case when $\Delta <0$, the absolute value of $\chi(n)$ decreases with increasing $n$.

Note that in Eqs.\ (\ref{chi(n)-3}) and (\ref{chi(n)-4}) we used ${\tilde g}= \sqrt{2}\, g$ and $\tilde{\tilde g}= \sqrt{3}\, g$. If a better approximation is used,
    \be \tilde g\approx \sqrt{2}\, g \, (1-\delta_{\rm q} /2\omega_{\rm q}^{\rm b}), \,\,\,
    \tilde {\tilde g}\approx \sqrt{3}\, g \, (1-\delta_{\rm q} /\omega_{\rm q}^{\rm b}), \,\,
    \label{g-tilde-tilde}\ee
then correction to the $\tilde g^2$-term in Eq.\ (\ref{omega-r-e-3lev}) creates an additional contribution $g^2\delta_{\rm q}/[\omega_{\rm q}^{\rm b}(\Delta-\delta_{\rm q})]$ to $\chi$, which for typical parameters is much larger than the second terms in Eqs.\ (\ref{chi(n)-3}) and (\ref{chi(n)-4}). The correction (\ref{g-tilde-tilde}) also yields an additional $n$-dependent contribution of
approximately $2g^4\delta_{\rm q}n/(\Delta^3\omega_{\rm q}^{\rm b})$ (assuming $\delta_{\rm q}\ll |\Delta |$) to both $\omega_{\rm r}^{|g\rangle}(n)$ and $\omega_{\rm r}^{|e\rangle}(n)$. These additional slopes are comparable to the slope of $\chi(n)$ in Eq.\ (\ref{chi(n)-4}) because $\Delta^2/(\delta_{\rm q}\omega_{\rm q}^{\rm b})$ is on the order of 1 for typical experimental parameters. However, the contribution to $\chi(n)$ from the correction (\ref{g-tilde-tilde}) is only $9 g^4\delta_{\rm q}^2n/(\Delta^4 \omega_{\rm q}^{\rm b})$ (assuming $\delta_{\rm q}\ll |\Delta |$), which is smaller than the last term in Eq.\ (\ref{chi(n)-4}) because $|\Delta |\ll\omega_{\rm q}^{\rm b}$. Therefore, the slope of $\chi(n)$ in Eq.\ (\ref{chi(n)-4}) is practically not affected by the correction (\ref{g-tilde-tilde}). Similarly, inaccuracy of our approximation of the fourth qubit level energy by $3\omega_{\rm q}^{\rm b}-3\delta_{\rm q}$ produces only a small correction to the slope of $\chi(n)$: the correction to the fourth-level energy is on the order of $\delta_{\rm q}^2/\omega_{\rm q}^{\rm b}$, and therefore the correction to $\chi (n)$ [via Eq.\ (\ref{fourth-level-correction})] is on the order of $g^4\delta_{\rm q}^2n/(\Delta^4 \omega_{\rm q}^{\rm b})$ for $\delta_{\rm q}\ll |\Delta |$.

Thus, we see that for a transmon or Xmon qubit, calculation of $\chi (0)$ requires at least three qubit levels to be taken into account, while the first correction due to $\chi (n)$ dependence ($n\ll n_{\rm crit}$) requires at least four qubit levels to be considered. In contrast, calculation of the Purcell rate $\Gamma (0)$ requires two qubit levels, while the first correction due to $\Gamma (n)$ dependence requires three qubit levels.

\subsection{Measurement error}

Now let us discuss the qubit measurement error caused by the qubit
relaxation due to Purcell effect.  To describe measurement, we will
use the dispersive approximation (\ref{H-disp}) with
$\chi=-g^2\delta_{\rm q}/\Delta^2$ [Eq.\ (\ref{chi-approx})] and
neglect the small difference between $\Delta$ defined in Eq.\
(\ref{Delta-def}) and $\omega_{\rm q}-\omega_{\rm r}$ defined via
Eq.\ (\ref{omega-q-r}). For the Purcell rate we will use the
simplest form $\Gamma = g^2\kappa/\Delta^2$. Both approximations
assume a sufficiently small number of photons, Eq.\
(\ref{n<<n-crit}).

Assuming that the qubit is either in the state $|g\rangle$ or
$|e\rangle$ (non-evolving), from Eq.\ (\ref{H-disp}) we see that the
effective resonator frequency is constant in time, $\omega_{\rm
r}\pm \chi$, where the upper sign is for the state $|e\rangle$ and
the lower sign is for $|g\rangle$. The corresponding resonator state
is a coherent state $|\alpha_{\pm}\rangle$, characterized by an
amplitude $\alpha_{\pm}(t)$ in the rotating frame based on the drive
frequency $\omega_{\rm d}$ [in the lab frame the resonator wave
function is $e^{-|\alpha_\pm|^2/2}\sum_n (\alpha_\pm e^{-i\omega_d
t})^n n^{-1/2} |n\rangle$ up to an overall phase]. The evolution of
the amplitude is
    \be\label{alpha-dot}
\dot \alpha_{\pm}= -i (\Delta_{\rm rd} \pm \chi)\, \alpha_{\pm}
-\frac{\kappa}{2}\, \alpha_{\pm} -i\varepsilon_{\rm r},
\end{equation}
where the upper sign everywhere is for the state $|e\rangle$,
    \be
    \Delta_{\rm rd} = \omega_{\rm r}-\omega_{\rm d},
    \ee
and complex $\varepsilon_{\rm r} (t)$ can in general depend on time to
describe a drive with changing amplitude and frequency. (Note a difference in notation compared with the main text: now $\Delta_{\rm rd}$ does not depend on the qubit state and the frequency shift $\pm \chi$ is added explicitly.) The steady
state for a steady drive, $\varepsilon_{\rm r}= {\rm const}$, is
    \be
    \alpha_\pm = \frac{-i\varepsilon_{\rm r}}
    {\kappa/2+i(\Delta_{\rm rd} \pm \chi)},
    \ee
so that the two coherent states are separated by
    \be
    \alpha_+ -\alpha_- = \frac{-2\varepsilon_{\rm r}\chi}
    {(\kappa/2+i\Delta_{\rm rd})^2+\chi^2},
    \ee
and this is the difference, which can be sensed by the homodyne
detection (it does not matter whether the microwave is transmitted or reflected from the resonator when we discuss the measurement in terms of $\alpha_{\pm}$).

In the case when $\Delta_{\rm rd}=0$ (so that $\omega_{\rm
d}=\omega_{\rm r}$), both states have the same average number of
photons, $\bar{n}=|\alpha_+|^2=|\alpha_-|^2=|\varepsilon_{\rm
r}|^2/(\kappa^2/4+\chi^2)$, and the absolute value of the state
separation is
    \be
    \delta \alpha \equiv | \alpha_+ -\alpha_-|= \frac{2 \sqrt{\bar{n}}}
    {\sqrt{(\kappa/2\chi)^2+1}}.
    \label{delta-alpha}\ee

Theoretically, the distinguishability of two coherent states separated by
$\delta\alpha$ is the same as the distinguishability of two Gaussians
with width (standard deviation) of 1/2 each, separated by
$\delta\alpha$. However, a measurement for the duration $t_{\rm m}$
increases the effective separation by a factor of $\sqrt{\kappa
t_{\rm m}}$, while imperfect quantum efficiency $\eta$ of the
measurement (mainly due to the amplifier noise) increases the
Gaussian width by $\eta^{-1/2}$ or, equivalently, decreases the
effective separation by $\eta^{-1/2}$. Therefore, we may think about the
distinguishability of two Gaussians with width 1/2 each, separated
by
    \be
    \delta \alpha_{\rm eff} = \sqrt{\eta \kappa t_{\rm m}} \,
    \delta\alpha ,
    \label{delta-alpha-eff}\ee
which gives the error probability due to state separation
    \be
    P_{\rm err}^{\rm sep}=\frac{1-{\rm Erf}(\delta\alpha_{\rm
    eff}/\sqrt{2})}{2} \approx
    \frac{\exp [-(\delta\alpha_{\rm eff})^2/2]}
    {\sqrt{2\pi}\,\delta\alpha_{\rm eff} } .
    \label{P-err-sep}\ee

The other contribution to the measurement error $P_{\rm err}$ for the state $|e\rangle$ comes
from the qubit energy relaxation with the rate $\Gamma+T_{1,\rm
int}^{-1}$ during the measurement time $t_{\rm m}$,
    \be
    P_{\rm err}=  P_{\rm err}^{\rm sep} + \frac{1}{2}\, t_{\rm m} (\Gamma+T_{1,\rm
int}^{-1}),
    \label{P-err}\ee
where the factor $1/2$ is because the relaxation moment is distributed practically uniformly within the measurement duration $t_{\rm m}$.
Since $P_{\rm err}^{\rm sep}$ decreases with time $t_{\rm m}$
exponentially, the main limitation for $P_{\rm err}$ comes from the
second term.

It is easy to find that $ P_{\rm err}^{\rm sep}=10^{-2}$ corresponds
to $\delta\alpha_{\rm eff}=2.3$, $ P_{\rm err}^{\rm sep}=10^{-3}$
corresponds to $\delta\alpha_{\rm eff}=3.1$, and $ P_{\rm err}^{\rm
sep}=10^{-4}$ corresponds to $\delta\alpha_{\rm eff}=3.7$. For an
estimate let us choose $\delta\alpha_{\rm eff}\agt 3$. Then from
Eq.\ (\ref{delta-alpha-eff}) $t_{\rm m}\agt
9/(\eta\kappa\,\delta\alpha^2)$, and therefore even neglecting
intrinsic relaxation $T_{1,\rm int}^{-1}$ in Eq.\ (\ref{P-err}), we
obtain the condition
    \be
    \Gamma \alt \frac{1}{4} P_{\rm err}\eta \kappa (\delta\alpha)^2.
    \ee
Now using $\Gamma=\kappa g^2/\Delta^2$ and assuming $\kappa \agt 2\, |\chi |$
in Eq.\ (\ref{delta-alpha}), so that $\delta\alpha \simeq 4\chi
\sqrt{\nbar}/\kappa$,
we rewrite this condition as
    \be
    \kappa \alt 2 \sqrt{P_{\rm err}\eta \nbar} \, \frac{|\chi|}{|g|}  \simeq  2 \sqrt{P_{\rm err}\eta \nbar}\,
    \frac{|g\delta_{\rm q}|}{|\Delta|},
    \ee
where for the second expression we used $\chi \simeq -g^2\delta_{\rm q}/\Delta^2$.
Since the measurement time $t_{\rm m}$ should be at least few times
longer than $\kappa^{-1}$, we obtain
    \be
t_{\rm m} >\frac{4}{\kappa} \agt \frac{2}{\sqrt{P_{\rm err}\eta
\nbar}} \, \frac{|\Delta|}{|g\delta_{\rm q}|} =\frac{4/\delta_{\rm q}}{\sqrt{P_{\rm err}\eta}\sqrt{\nbar/n_{\rm crit}}}.
    \label{tm-lim}\ee

These estimates show that the Purcell effect requires a sufficiently long measurement time when we desire a small measurement error $P_{\rm err}$. The limitation is not severe, but it is still inconsistent with a fast accurate measurement needed for quantum error correction. As an example, from Eq.\ (\ref{tm-lim}) we see that for $\delta_{\rm q}/2\pi\simeq 200\, {\rm MHz}$, $P_{\rm err}\simeq 10^{-3}$, $\eta \simeq 0.3$ and $\nbar \simeq n_{\rm crit}/4$ we need $t_m >400\,{\rm ns}$ (this in turn would require a very long $T_{1, \rm int}$). Also, the detuning should be sufficiently large,
    \be
|\Delta/g|>\sqrt{\kappa t_{\rm m}/2P_{\rm err}}>\sqrt{2/P_{\rm err}},
    \ee
as directly follows from $\Gamma =\kappa (g/\Delta)^2$ and $\kappa t_{\rm m} >4$.

As a more detailed example, let us choose $g/2\pi \simeq 30\, {\rm MHz}$, $\Delta/2\pi \simeq -1.35\, {\rm GHz}$, $\delta_{\rm q}/2\pi\simeq 200\, {\rm MHz}$, $\kappa^{-1}\simeq 100 \, {\rm ns}$, $t_{\rm m}\simeq 400 \, {\rm ns}$, $\eta\simeq 0.3$, and $\nbar\simeq n_{\rm crit}/4\simeq 125$. Then for the excited qubit state the Purcell decay brings the error contribution $t_{\rm m}\Gamma /2\simeq 10^{-3}$; the dispersive coupling is $\chi /2\pi \simeq -0.1 \, {\rm MHz}$, so $\delta \alpha \simeq 0.25\sqrt{\nbar}\simeq 2.8$ and $\delta\alpha_{\rm eff}\simeq 3$, so the separation error is also about $10^{-3}$ [actually slightly larger since $|\chi|$ decreases with $\nbar$ for negative $\Delta$ -- see Eq.\ (\ref{chi-slope})].

Thus, we see that even with the qubit decay due to the Purcell effect, it is possible to measure a qubit with a low measurement error, but this requires a relatively long time and large resonator-qubit detuning. Note that we also had to assume a large number of photons in the resonator, which can lead to detrimental effects (neglected in our analysis) such as dressed dephasing \cite{Boi08,Boi09} and various imperfections related to nonlinear dynamics. Suppression of the Purcell effect (using a Purcell filter or other means) allows us to significantly increase the ratio $|g/\Delta|$, thus increasing the coupling $\chi$ and correspondingly decreasing the measurement time for the same measurement error.


\begin{thebibliography}{99}

\bibitem{N-C-book} M. A. Nielsen and I. L. Chuang, {\it Quantum
Computation and Quantum Information} (Cambridge University Press,
Cambridge, 2000).

\bibitem{Barends-14} R. Barends, J. Kelly, A. Megrant, A. Veitia,
D. Sank, E. Jeffrey, T. C. White, J. Mutus, A. G. Fowler, B.
Campbell, Y. Chen, Z. Chen, B. Chiaro, A. Dunsworth, C. Neill, P.
O'Malley, P. Roushan, A. Vainsencher, J. Wenner, A. N. Korotkov, A.
N. Cleland, and J. M. Martinis, Nature {\bf 508}, 500 (2014).

\bibitem{Chow-14} J. M. Chow, J. M. Gambetta, E. Magesan, D. W. Abraham,
A. W. Cross, B. R. Johnson, N. A. Masluk, C. A. Ryan, J. A. Smolin,
S. J. Srinivasan, J. Srikanth, and M. Steffen, Nature Comm. {\bf 5},
4015 (2014).

\bibitem{Weber-14} S. J. Weber, A. Chantasri, J. Dressel, A. N. Jordan,
K. W. Murch, and I. Siddiqi, Nature {\bf 511}, 570 (2014).

\bibitem{Sun-14} L. Sun, A. Petrenko, Z. Leghtas, B. Vlastakis,
G. Kirchmair, K. M. Sliwa, A. Narla, M. Hatridge, S. Shankar, J.
Blumoff, L. Frunzio, M. Mirrahimi, M. H. Devoret, and R. J.
Schoelkopf, Nature {\bf 511}, 7510 (2014).

\bibitem{Riste-14} D. Riste, M. Dukalski, C. A. Watson, G. de Lange,
M. J. Tiggelman, Y. M. Blanter, K. W. Lehnert, R. N. Schouten, and
L. DiCarlo, Nature {\bf 502}, 350 (2013).

\bibitem{Stern-14} M. Stern, G. Catelani, Y. Kubo, C. Grezes,
A. Bienfait, D. Vion, D. Esteve, and P. Bertet, Phys. Rev. Lett.
{\bf 113}, 123601 (2014).

\bibitem{Mlynek-14} J. A. Mlynek, A. A. Abdumalikov, C. Eichler, and
A. Wallraff, Nature Comm. {\bf 5}, 5186 (2014).

\bibitem{Lin-14} Z. R. Lin, K. Inomata, K. Koshino, W. D. Oliver, Y. Nakamura, J.
S. Tsai, and T. Yamamoto, Nature Comm. 5, 4480 (2014).

\bibitem{Kelly-15}  J. Kelly, R. Barends, A. G. Fowler, A. Megrant,
E. Jeffrey, T. C. White, D. Sank, J. Y. Mutus, B. Campbell, Yu Chen,
Z. Chen, B. Chiaro, A. Dunsworth, I.-C. Hoi, C. Neill, P. J. J.
O'Malley, C. Quintana, P. Roushan, A. Vainsencher, J. Wenner, A. N.
Cleland, and J. M. Martinis, Nature {\bf 519}, 66 (2015).



\bibitem{Blais-04} A. Blais, R. S. Huang, A. Wallraff,
S. M. Girvin, and R. J. Schoelkopf, Phys. Rev. A {\bf 69}, 062320
(2004).

\bibitem{Wallraff-04} A. Wallraff, D. I. Schuster, A. Blais, L. Frunzio,
R. S. Huang, J. Majer, S. Kumar, S. M. Girvin, and R. J. Schoelkopf,
Nature (London) {\bf 431}, 162 (2004).

\bibitem{Esteve-86} D. Esteve, M. H. Devoret, and J. M. Martinis,
Phys. Rev. B {\bf 34}, 158 (1986).

\bibitem{Purcell-46} E. M. Purcell, Phys. Rev. {\bf 69}, 681 (1946).


\bibitem{Haroche-book} H. Haroche and J.-M. Raimond, \textit{Exploring the Quantum:Atoms, Cavities, and Photons}
(Oxford University Press, Oxford, 2006).


\bibitem{Red10} M. D. Reed, B. R. Johnson, A. A. Houck, L. DiCarlo,
J. M. Chow, D. I. Schuster, L. Frunzio, and R. J. Schoelkopf, Appl.
Phys. Lett. \textbf{96}, 203110 (2010).

\bibitem{Jeffrey-14} E. Jeffrey, D. Sank, J.Y. Mutus, T.C. White,
J. Kelly, R. Barends, Y. Chen, Z. Chen, B. Chiaro, A. Dunsworth, A.
Megrant, P.J.J. O'Malley, C. Neill, P. Roushan, A. Vainsencher, J.
Wenner, A.N. Cleland, J. M. Martinis, Phys. Rev. Lett. \textbf{112},
190504 (2014).

\bibitem{Bronn-APS-15} N. Bronn, A. Corcoles, J. Hertzberg, S. Srinivasan, J. Chow, J. Gambetta, M. Steffen, Y. Liu, and  A. Houck, Bull. Am. Phys. Soc.
    {\bf 60}, S39.6 (2015).


\bibitem{Bronn-15} N. T. Bronn, E. Magesan, N. A. Masluk, J. M. Chow, J. M. Gambetta and M. Steffen, arXiv:1504.04353.

\bibitem{Gam11} J. M. Gambetta, A. A. Houck, and A. Blais, Phys. Rev.
Lett. \textbf{106}, 030502 (2011).

\bibitem{Srinivasan-11} S. J. Srinivasan, A. J. Hoffman, J. M. Gambetta,
 and A. A. Houck, Phys. Rev. Lett. {\bf 106}, 083601 (2011).


\bibitem{Set13} E. A. Sete, A. Galiautdinov, E. Mlinar,
J. M. Martinis, and A. N. Korotkov, Phys. Rev. Lett. \textbf{110},
210501 (2013).


\bibitem{Set14} E. A. Sete, J. M. Gambetta, and A.N. Korotkov, Phys. Rev. B \textbf{89}, 104516 (2014).

\bibitem{Koch-07} J. Koch, T. M. Yu, J. Gambetta, A. A. Houck, D. I.
Schuster, J. Majer, A. Blais, M. H. Devoret, S. M. Girvin, and R. J.
Schoelkopf, Phys. Rev. A {\bf 76}, 042319 (2007).

\bibitem{Boisson-10} M. Boissonneault, J. M. Gambetta, and A. Blais, Phys. Rev. Lett. {\bf 105}, 100504 (2010).

\bibitem{Allen-Eberly-book} L. Allen and J. H. Eberly, {\it Optical Resonance and Two-Level Atoms} (Dover, Mineola, NY, 1997).

\bibitem{RWA-note} The rotating wave approximation neglects the counter-rotating terms. For driven classical oscillators [Eqs.\ (\ref{alpha-dot-1}) and (\ref{beta-dot-1})] it requires that $|\Delta_{\rm rd}|$, $|\Delta_{\rm fd}|$, $|\mathcal{G}|$, and $|\dot{\varepsilon_{\rm r}}/\varepsilon |$ are small compared to $\omega_{\rm r}$.

\bibitem{Scully-book}  M. O. Scully and M. S. Zubairy, {\it Quantum Optics} (Cambridge Univ. Press, Cambridge, 1997).


\bibitem{Kor-13} A. N. Korotkov, arXiv:1309.6405, Appendix B.

\bibitem{Pie07} P. Meystre and M. Surgent III, \textit{Elements of
Quantum Optics} (Springer-Verlag, Berlin, 2007).

\bibitem{Car93} H. J. Carmichael, Phys. Rev. Lett. \textbf{70}, 2273 (1993).

\bibitem{Boi08} M. Boissonneault, J. M. Gambetta, and A. Blais, Phys. Rev. A
77, 060305 (2008).
\bibitem{Boi09} M. Boissonneault, J. M. Gambetta, and A. Blais, Phys. Rev. A
79, 013819 (2009).


\end{thebibliography}
\end{document}